\documentclass{arxpub} % uncomment this for numbers citation

\usepackage[colorlinks = true,linkcolor = blue,urlcolor  = blue,citecolor = blue,anchorcolor = blue]{hyperref}

\usepackage[margin=1.0in]{geometry}
\usepackage{parskip} 

\usepackage{amsmath}
\usepackage{amssymb}
\usepackage{bbold}
\usepackage{mathtools}
\usepackage{enumerate} % allows for roman numeral lists
\usepackage{xcolor}
\usepackage{graphicx}
\usepackage{subfig}
\usepackage{float}
\usepackage[ruled]{algorithm2e}
\usepackage{ulem}[normalem]
\usepackage{cancel}
\usepackage[numbers,sort&compress]{natbib}

\authornames{Marion, Mathews and Schmidler}
\shorttitle{Path Selection for SMC}

\numberwithin{equation}{section}
\theoremstyle{plain}

\newcommand{\asref}[1]{AS\ref{#1}}

\DeclarePairedDelimiter{\ceil}{\lceil}{\rceil}
 
\newcommand{\bigO}{\mathcal{O}}
\DeclareMathOperator{\E}{E}
\newcommand{\Var}{\text{Var}}
\newcommand{\Lp}[3]{\norm{#1/#2}_{L_{#3}(#2)}^{#3}}
\newcommand{\iidsim}{\stackrel{\mbox{\tiny{iid}}}{\sim}}

\newcommand{\A}[1]{\textbf{A}_{#1}}
\newcommand{\B}[1]{\textbf{B}_{#1}}
\newcommand{\C}[1]{\textbf{C}_{#1}}
\newcommand{\D}[1]{\textbf{D}_{#1}}
\newcommand{\BB}[1]{\bar{\textbf{B}}_{#1}}
\newcommand{\DD}[1]{\bar{\textbf{D}}_{#1}}

\newcommand{\V}{\mathcal{V}}
\newcommand{\X}{\mathcal{X}}

\newcommand{\norm}[1]{\left\lVert#1\right\rVert}

\newcommand{\OS}[1]{\mathcal{O}^*(#1)}

\begin{document}

\title{Finite sample bounds for sequential Monte Carlo and adaptive path selection using the $L_2$ norm.}

\authorone[Berry Consultants]{Joe Marion}
\authortwo[Duke University]{Joseph Mathews} 
\authorthree[Duke University]{Scott C. Schmidler} 

%Please use the following format for addresses and emails. The APT office will sort this out after you submit your files.
\addressone{Austin, TX 78746, USA} % Your postal address goes here.
\emailone{joseph@berryconsultants.net} %Authors email goes here.

\begin{abstract}
We prove a bound on the finite sample error of sequential Monte Carlo (SMC) on static spaces using the $L_2$ distance between interpolating distributions and the mixing times of Markov kernels. This result is unique in that it is the first finite sample convergence result for SMC that does not require an upper bound on the importance weights.  Using this bound we show that careful selection of the interpolating distributions can lead to substantial improvements in the computational complexity of the algorithm. This result also justifies the adaptive selection of SMC distributions using the relative effective sample size commonly used in the literature, and we establish conditions guaranteeing the approximation accuracy of the adaptive SMC approach. We show that the commonly used data tempering approach fails to satisfy these conditions, and introduce a modified data tempering algorithm under which our guarantees do hold. We then demonstrate empirically that this procedure provides nearly-optimal sequences of distributions in an automatic fashion for realistic examples.
\end{abstract}

%\begin{keyword}
%\kwd{Sequential Monte Carlo}
%\kwd{computational complexity}
%\kwd{path selection}
%\kwd{Bayesian computation}
%\end{keyword}

\newpage
\section{Introduction}
Sequential Monte Carlo (SMC) is a sampling method that moves particles drawn from an initial distribution $\mu_0$ to a target distribution $\pi $ via a sequence of interpolating distributions $\mu_0,\ldots,\mu_S=\pi$.  Choosing an appropriate sequence of distributions, which we refer to as a {\it path}~\citep{GelmanMeng, lin2013}, is critical to obtaining an efficient SMC sampler. Common path selection approaches for static (fixed dimension) SMC problems include batch processing of data~\citep{chopin2002sequential}, tempering with pre-determined schedules~\citep{del2006sequential, neal2001annealed,zhou2016toward}, and tempering with adaptively chosen temperatures~\citep{jasra2015levy, zhou2016toward, lin2013}. Comparison of paths is generally limited to simulation studies; the theoretical SMC literature treats the sequence of interpolating distributions as given and does not generally account for the impact of different path choices, nor the effects of automated path selection techniques~\citep{whiteley2012sequential,schweizer2012non, jasra2015error, chopin2004central, douc2008, DelMoral2004Feynman}. 

In the first part of this paper, we directly relate the computational complexity of obtaining a bounded-error SMC estimator to the selection of interpolating distributions.  More formally, we demonstrate conditions under which SMC provides a \textit{randomized approximation scheme} for estimating expectations of $\pi$. The bound presented here improves on the results of~\cite{Marion2018}, relaxing the assumption of bounded density ratios, requiring only a bound on the $L_2$ distance between adjacent distributions instead. This allows us to explicitly relate the distributions in the selected path to the error in the resulting estimator and the computational complexity of the algorithm. This in turn allows us to identify sequences of interpolating distributions (paths) that lead to substantial improvements in efficiency. Unlike other finite sample results for SMC in the literature \citep{Marion2018, whiteley2012sequential, schweizer2012non}, it also enables us to establish the convergence of SMC in situations where the importance sampling weights are unbounded.  We apply this approach to quantify improvements obtained by alternate path selection on two examples. The first is a spherical Gaussian target distribution, where we show that a two-dimensional path using geometric mixtures that also alters the precision has superior complexity to a one-dimensional path using only geometric mixture. The second example considers general log-concave target distributions. We show that combining the path from~\cite{LovaszON3} with the sampling algorithm from~\cite{WuMALA} provides an upper bound for SMC that obtains state of the art complexity for this problem.

In practice, pre-specifying a sequence of distributions that efficiently controls the $L_2$ distance between steps during the application of SMC to new problems may be difficult. The next section of the paper analyzes a practical scheme for adaptively choosing a sequence of distributions so that the $L_2$ distance between steps is provably controlled when weights are bounded.  This is accomplished through monitoring the relative effective sample size (RESS). Adaptive path selection using the RESS is well known to the SMC community~\citep{jasra2015levy, lin2013, zhou2016toward}; we provide conditions under which the RESS can be shown to estimate the $L_2$ distance between steps with high accuracy, providing rigorous support for its use in choosing an SMC path. We then provide error bounds in this adaptive setting, giving  conditions under which SMC using adaptive path selection remains a randomized approximation scheme. We conclude by demonstrating the empirical performance of this adaptive algorithm on two examples. First, a mean field Ising model, where we show that adaptive SMC using tempered distributions finds nearly optimal sequences of interpolating distributions. The second is a Bayesian linear regression problem, where we demonstrate that the commonly-used data-tempering approach to path selection may result in steps with unexpectedly large $L_2$ distances, causing significant instability in the resulting estimator. To address this issue, we introduce a hybrid path construction that combines the computational advantages of data-tempering with the stability of traditional tempering and yields guarantees on approximation error using our bounds.

\section{Sequential Monte Carlo error bounds}
\label{sec:L2Bound}
In this section, we present the main results of the paper.  Before doing so, we introduce some notation and describe the SMC algorithm studied in this paper.
 
\subsection{Notation} Let $\big(\X, \mathcal{B}, \lambda)$ be a probability space. Define $\mathcal{P}$ the set of probability measures on $\X$ that are absolutely continuous with respect to $\lambda$ and $\mathcal{F}$ the set of measurable functions $f:\X \rightarrow\mathbb{R}$. We denote the expectation of $f\in\mathcal{F}$ with respect to a measure $\mu$ by $\mu(f) := \int f(x)\mu(dx)= \E(f)$. 
% Each measure acts on functions $f\in\mathcal{F}$ \textit{from the left}\scsfootnote{Not clear what this means here. Usually might write $\mu f$ to match operator notation for $K$ (that is, $\mu K$ and $Kf$ for $K$ operating on measures from the left and $L_2$ functions from the right).  Also: this would be a good place to define the empirical measure notation if we want to use that (see comments around statement of Thm 2.1 below).} by $\mu(f) = \int f(x)\mu(dx)= \E(f)$. 

The convergence results in this paper depend on the $L_2$ distance between interpolating distributions used in SMC, and the mixing times of the corresponding Markov kernels.

\paragraph{$L_2$ distance:}  For $\mu,\eta \in \mathcal{P}$ define the \textit{$L_2$ distance from $\mu$ to $\eta$} by the $L_2(\mu)$ norm of $\eta/\mu$:
\begin{equation}
\|\eta/\mu\|_{L_2(\mu)}^2 = \int\Big(\frac{d\eta}{d\mu}(x)\Big)^2 \mu(dx)
\end{equation}
% \scsfootnotetext{This should be either $\frac{\eta(x)}{\mu(x)}$ (if we are assuming densities), or more generally 
% $\frac{d\eta}{d\mu}(x)$}
%
Although not a true metric, the $L_2$ distance provides a measure of the discrepancy between $\mu$ and $\eta$; subtracting one yields the traditional $\chi^2$  divergence from $\mu$ to $\eta$, which gives the variance of the importance sampling weights. As discussed in Section~\ref{sec:adaptive}, the $L_2$ distance is also related to the relative effective sample size (RESS), a quantity that is used to assess the degeneracy of the particle system.

\paragraph{Mixing times:}
We say that a measure $\nu\in\mathcal{P}$ is $\omega$-warm with respect to $\mu$ if $\omega =  \sup_{B\in\mathcal{B}} \;\nu(B) / \mu(B)$~\citep{Lovasz04, Vempala05}. Let $\mathcal{P}_\omega (\mu)$ be the set of all such measures.  For an ergodic Markov kernel $K:\X\times\mathcal{B}\rightarrow[0,1]$ with limiting distribution $\mu$, define the $\omega$-warm mixing time of $K$ by $\tau_K(\epsilon, \omega) = \min\; \big\{t : \sup_{\nu \in \mathcal{P}_\omega(\mu)} \|\nu K^t - \mu\|_{\text{TV}}\leq \epsilon\big\}$, where 
$\|\cdot\|_{\text{TV}}$ 
denotes %the 
total variation norm.  

\paragraph{Randomized approximation:} An algorithm for producing a Monte Carlo approximation $\hat{f}$ of $\pi(f)$ 
%$\hat{\pi}f$ of $\pi f$
is a \textit{randomized approximation scheme} if, for any user-specified $\epsilon>0$ and $\delta \in (0,1]$, it guarantees %\scsfootnote{We haven't defined $\hat{\pi}$. Change to $\abs{\hat{f} - \pi(f)}$. Or if we need this notation later, 
% put $\hat{f} =\hat{\pi}f=\hat{\mu}_Sf = \frac{1}{N} \ldots$ in the definition of $\hat{f}$ above, and define the empirical measure $\hat{\pi}$ somewhere, and make sure the notation $\mu f$ is defined (instead of $\mu(f)$) so that $\hat{\pi}f$ and $\pi f$ are well defined here.}
%$|\hat{\pi}f - \pi f|<\epsilon$
$|\hat{f} - \pi(f)|<\epsilon$
with probability at least $1-\delta$~\citep{Motwani1995RandomizedA}. For ease of presentation we establish this for $\delta=$ 1/4, but this is easily improved to arbitrary $\delta>0$ at the cost of an additional factor of $\mathcal{O}\big(\log(1/\delta)\big)$ using the median approach (see Lemma 6.1 of \cite{JERRUM1986}). 

%\paragraph{$L_2$ distance} 

\subsection{Sequential Monte Carlo}
\label{Sec:SMCIntro}
In this paper we study the following SMC algorithm. Before sampling, the user specifies a path $\mu_0,\ldots,\mu_S$ where $\mu_s\in\mathcal{P}$ and $\mu_{s-1} \ll \mu_s$. We abuse notation and write the density $\mu_s(x) = q_s(x)/z_s$ where $q_s(x)$ is a known, unnormalized density.  The algorithm is initialized by drawing $N$ samples $X_0^{1:N} = X_0^{1},\ldots, X_0^N$ independently from $\mu_0$, then proceeds in $S$ steps. At the beginning of step $s$, each particle is assigned an importance sampling weight $w_s(X_{s-1}^n)=q_s(X_{s-1}^n)/q_{s-1}(X_{s-1}^n)$. Then, a new set of particles $\tilde{X}_s^{1:N}$ is drawn with replacement from the current particles according to the weights (multinomial resampling); i.e. a copy of $X_{s-1}^n$ is drawn with probability proportional to $w_s(X_{s-1}^n)$. Finally, each resampled particle evolves independently according to a Markov kernel $K_s$ with stationary distribution $\mu_s$, resulting in a new set of particles $X_s^{n} \sim K_s^t\big(\tilde X_s^n, \;\cdot \;\big)$. Following step $S$ of the algorithm $\pi (f)$ is estimated using the particle average $\hat f := \frac{1}{N} \sum_{n=1}^N f\big(X_{S}^{n}\big)$. Detailed descriptions of this SMC algorithm can be found in~\cite{del2006sequential,chopin2002sequential,Marion2018}.

\subsection{Adaptive path selection}
Specifiying an effective  path for SMC can be difficult in practice. In Section~\ref{sec:adaptive} we study a variation of the above SMC algorithm in which the path need not be specified in advance, but can be chosen adaptively.
Adaptive SMC algorithms 
%address this problem by using
use information from the   particle system at each step to  dynamically select the next distribution from a set of candidates.  A common adaptive approach is to choose the next distribution by comparing the relative effective sample size (RESS) for different possible candidates~\citep{jasra2015levy, zhou2016toward, lin2013}.  The RESS %when 
moving from $\mu_{s-1}$ to $\mu_s$ is defined by:
\begin{equation}
\label{eq:selection1}
%   {E}_s = \frac{\Big(N^{-1}\sum_{s=1}^N w_s\big(x^{n}_{s-1} \big)\Big)^2}{N^{-1}\sum_{s=1}^N w_s\big(x_{s-1}^{n}\big)^2}
{E}_s = \big(N^{-1}\sum_{n=1}^N w_s(x^{n}_{s-1})\big)^2 
\,/ \,\big( N^{-1}\sum_{n=1}^N w_s(x_{s-1}^{n})^2\big)
\end{equation}
and is interpreted as the ratio of the (estimated asymptotic) variance of the SMC estimator $\hat{f}_s$ at step $s$ to that obtained by independent sampling. When ${E}_s$ is small the particle system is said to be %described as 
\textit{degenerate}. 

\subsection{Error bounds for SMC}
The results in this paper provide bounds on the approximation error of SMC.  Our approach is based on bounding the $L_2$ distance between successive distributions in the path. Our first main result is Theorem~\ref{thm:total_bound}, which establishes conditions under which SMC serves as a randomized approximation scheme. 
Theorem~\ref{thm:total_bound} requires the following assumptions:
\begin{assumption}\label{as:1}
   $ \|\mu_s/\mu_{s-1}\|_{L_2(\mu_{s-1})}^2 \leq \mathcal{E}^{-1} < \infty $ for $s=1,\ldots,S.$
\end{assumption}
\begin{assumption}\label{as:2}
    $K_s$ has limiting distribution $\mu_s$ with mixing time $\tau_s(\epsilon, \omega)$.
\end{assumption}
\noindent Here $\mathcal{E}^{-1}$ bounds the maximal $L_2$ distance between adjacent distributions.  Under these conditions, we have the following result:
% Thing
\begin{theorem}[Error bound for SMC]\label{thm:total_bound}
\hfill \break
Assume~\asref{as:1} and \asref{as:2}. Fix $\epsilon > 0$ and sample $X_0^{1:N}$ independently from $\mu_0$. Let
\begin{itemize}
    \item[1.] $N\geq \log\big(128S\big)\cdot \max\big\{\frac{18}{\mathcal{E}}, \frac{1}{2\epsilon^2}\big\}$
    \item[2.] $t \geq \max_s\; \tau_s \big( \frac{1}{8NS},\; 2\big).$
\end{itemize}
Then for any $f\in\mathcal{F}$ with $|f|\leq 1$, we have
%
%\scsfootnote{Make the notation here consistent with whatever was chosen above.}
$\big|\hat{f} - \pi(f)\big| \leq \epsilon$
with probability at least $3/4$.
\end{theorem}
The proof of Theorem~\ref{thm:total_bound} is given in Appendix~\ref{sec:thm1} and closely follows the proof of Theorem 1 in~\cite{Marion2018}. The key difference is the use of Bernstein's inequality to ensure concentration of the weights, which replaces Lemma 4 of~\cite{Marion2018} and results in a modified one-step induction condition yielding Theorem~\ref{thm:total_bound} above. In addition, \asref{as:1} above replaces \asref{as:1} of~\cite{Marion2018}, which requires an upper bound $W$ on the weights and a lower bound $Z$ on the ratios of normalizing constants. When such bounds are available we immediately have  $\mathcal{E}^{-1}\leq W^2 Z^2$ to apply Theorem~\ref{thm:total_bound}. However, requiring a bound on $\mathcal{E}^{-1}$ instead has several advantages. First, the assumption of bounded weights restricts the sequences of interpolating distributions that can be considered and is frequently violated in applications. (Despite this, it is commonly assumed in theoretical results for both asymptotic and finite sample convergence of SMC).  Second, assumption~\asref{as:1} enables us to compare our resulting SMC bounds directly to bounds for MCMC, as explored in Section~\ref{sec:path_examples}.

\subsection{Error bounds for adaptive-path SMC}
Our second main result extends the SMC error bound of Theorem~\ref{thm:total_bound} to the adaptive-path setting.  Section~\ref{sec:adaptive_algorithm} describes an adaptive path SMC algorithm that uses 
RESS to dynamically choose the next distribution at each step. 
Theorem~\ref{thm:adaptive_total_bound} below gives conditions under which this adaptive algorithm provably constitutes a randomized approximation scheme.

Before stating this result,  we note that the RESS can also be interpreted as a sampling estimate of the (inverse) $L_2$ distance, since $1/{E}_s \rightarrow \|\mu_s/\mu_{s-1}\|_{L_2( \mu_{s-1})}^2$ as $n\rightarrow\infty$. Therefore when this RESS estimate is sufficiently accurate, it can be used to choose a path with bounded $L_2$ distance with high probability by ensuring ${E}_s>\mathcal{E}$ holds at each step with sufficiently high probability. In particular, for the adaptive-path SMC algorithm described in Section~\ref{sec:adaptive_algorithm}, we require the following assumptions:
\begin{assumption}\label{as:3}
   The weights are bounded for each possible step: 
   $\sup_{\mu \in \V,\, \nu \in \V(\mu),\, x\in\X} w_{\mu, \nu}(x) \leq 1.$
\end{assumption}
\begin{assumption}\label{as:4}
     The smallest step from every node has bounded $L_2$ norm: 
     $\sup_{\mu_s \in \V} \Lp{\nu_{1,s}}{\mu_s}{2} \leq 2 \mathcal{E}^{-1}$
\end{assumption}
\noindent \asref{as:3} is needed to provide conditions under which the $L_2$ distance can be accurately estimated. \asref{as:4}, discussed at length at the end of Section~\ref{sec:path_assumptions}, ensures that the algorithm does not terminate prematurely due to a lack of viable candidate distributions. In addition, the specification of the adaptive path algorithm in 
Section~\ref{sec:adaptive_algorithm} implicitly imposes further restrictions, which we state here as an additional assumption for clarity:
\begin{assumption}\label{as:5}
     Every path is finite and terminates in $\pi$, and the number candidate distributions for any step is bounded: $\sup_{\mu \in \V} | \V\left(\mu\right)| \leq M.$
\end{assumption}
%
%\moved{The requirement that the number of candidates is bounded is necessary to control the error of each ${E}_{s,m}$, which requires coupling and concentration inequalities for each possible candidate. This requirement is often satisfied in practice, even when the number of candidates is theoretically larger.  For example, while the geometric mixture or tempering is often described as using continuous temperature ladder, in practice adaptive algorithms select candidates using search over a finite number of candidates.} 

Under these conditions, we obtain the following result:
\begin{theorem}[Error bound for Adaptive Step-Selection SMC]\label{thm:adaptive_total_bound} Choose  $\mathcal{E}\in\left(0, 1\right)$ and $\mathcal{C} \in \left(0, 5/6\right)$ and assume~\asref{as:3}-\asref{as:5}.  For $s = 0,1,\ldots,S$ set
\begin{equation}
\label{Eqn:AdaptiveSampleSize}
N_s = \max 
\begin{cases}
%36 \log \Big(20 M\cdot \left(1+s^2\right)\Big) \cdot \mathcal{E}^{-1}\\ 
36 \cdot \gamma(s) \cdot \mathcal{E}^{-1}\\ 
%25/2 \log \Big(40 M\cdot \left(1+s^2\right)\Big) \cdot \mathcal{C}^{-2}\\
25/2 \cdot (\gamma(s) + \log(2)) \cdot \mathcal{C}^{-2}\\
%1/2 \log \Big(20M \cdot \left(1+s^2\right)\Big) \cdot \epsilon^{-2}\\
1/2 \cdot \gamma(s) \cdot \epsilon^{-2}
\end{cases}
\end{equation}
for $\gamma(s) = \log(20M(1+s^2))$. Define $\tau\left(\;\cdot\;, 2\right) = \sup_{\mu \in \V} \tau_\mu\left(\;\cdot\;, 2\right)$ and for $s\geq1$ set
\begin{equation*}
    t_s \geq \tau
    %\left( \frac{1}{ 16 s^ 2N_s},\; 2\right). 
    \big( (16 s^ 2N_s)^{-1},\; 2\big).
\end{equation*}
Fix $\epsilon > 0$ and draw $X_0^{1:N_0}\iidsim\mu_0$. Then for any $f\in\mathcal{F}$ with $|f|\leq 1$ the adaptive SMC algorithm ensures $\big|\hat f - \pi \left(f\right)\big| \leq \epsilon$ with probability at least $3/4$.
\end{theorem}
The proof of Theorem~\ref{thm:adaptive_total_bound} is given in Appendix~\ref{sec:thm2_proof}.  The argument is similar to that of Theorem~\ref{thm:total_bound}, but requires modifications to address the difficulties arising from the adaptive nature of the algorithm. First, at each step of the algorithm, the next interpolating distribution is random since it is chosen from based on the realization of the particle system; hence we require that the algorithm  
select each next step to guarantee a bound such as \asref{as:1} holds with high probability. Second, this  randomness in the choice of interpolating distributions means that the total number of steps $S$ in the selected path is also random.  We show that by gradually increasing the size of the particle system $N_s$ and the number of Markov transitions $t_s$ at each step of the algorithm, this problem can be overcome. Comparing Theorems~\ref{thm:total_bound} and~\ref{thm:adaptive_total_bound}, we see that surprisingly, the modifications that accommodate the randomness in $S$ do not increase the computational complexity of the algorithm relative to the performance of non-adaptive SMC on the same set of interpolating distributions chosen in advance.

To our knowledge Theorem~\ref{thm:adaptive_total_bound} provides the first error bounds for
%\scsfootnote{As I recall, our assumptions are what allow us to guarantee convergence.  So maybe we should call this the first ``error bounds for" rather than ``proof of convergence for"?} 
for SMC with adaptively chosen sequences of distributions.  Previously~\cite{Durham2014} and~\cite{zhou2016toward} employed a two stage approach to ensure that a central limit theorem held in the adaptive setting, first running adaptive SMC algorithm to select a path, followed by a non-adaptive SMC run on the selected path to estimate expectations under $\pi$. Theorem~\ref{thm:adaptive_total_bound} shows that this two stage procedure is unnecessary. In addition, the two stage procedure provides no guarantees about the finite sample properties of the adaptively chosen path. Theorem~\ref{thm:adaptive_total_bound} may also be seen as validating the use of the RESS for selecting distributions. Other step selection approaches have been considered~\citep{Martino2016, Nguyen2016}; however, these methods currently lack theoretical support.

We can make no claims at this point about near-optimality of the selected path's length.
%in terms of path length. 
We explore this issue empirically in Section~\ref{sec:empirical}.  Finally, the additional requirement of the lower bound $\mathcal{C}$ (described in \eqref{eq:selection}) may be unnecessarily restrictive in practice, but it is unclear at this time if selecting distributions without this condition (or a similar requirement) is sufficient. Theorem~\ref{thm:another_adaptive_total_bound} in the appendix provides an alternate result which does not require this condition, but requires a bound on the $L_4$ distance instead.

\section{Path selection and complexity}
\label{sec:path_examples}
The finite sample bounds given in Theorem~\ref{thm:total_bound} facilitate explicit comparison between algorithms. In this section, we compare the bounds on the computational  complexity of SMC obtained from Theorem~\ref{thm:total_bound} with available bounds for an alternative sampling algorithm (MCMC), highlighting the advantages of each approach. Complexity is given in total number of Markov kernel transitions required to approximate $\pi f$. Suppose that $K_1,\ldots,K_S$ are geometrically ergodic and reversible with respective spectral gaps $\rho_1,\ldots,\rho_S \in (0,1)$. Then the number of transitions according to $K_S$ required to sample approximately from $\pi$ using a Markov chain starting with a draw from $\mu_0$ is~\citep{roberts1997geometric,roberts2001}
\begin{equation*}    %\bigO\bigg(\frac{\log\|\mu_0/\pi\|_{L_2(\pi)}}{\rho_S}\bigg).
\bigO\big(\log\|\mu_0/\pi\|_{L_2(\pi)}/\rho_S\big).
\end{equation*}
In comparison, Theorem~\ref{thm:total_bound} gives the following complexity bound for SMC
\begin{equation*}
%    \bigO\bigg(\frac{S/\mathcal{E}\cdot \log^2\big(S/\mathcal{E}\big)}{\rho^*} \bigg)
\bigO\big(S/\mathcal{E}\cdot \log^2(S/\mathcal{E})/\rho^*\big)
\end{equation*}
where $\rho^*=\min_s \rho_s$. When the spectral gaps of the Markov kernels $K_s$ are of the same order, or if $\rho^* = \rho_S$ (i.e. the lowest temperature is the slowest mixing), the two bounds differ primarily by the cost of moving from the initial distribution to the target distribution. For MCMC, this factor is  $\log\|\mu_0/\pi\|_{L_2(\pi)}$, whereas for SMC this factor is an upper bound on $S/\mathcal{E} \geq S\cdot\max_s  \|\mu_s/\mu_{s-1}\|_{L_2(\mu_{s-1})}^2$, which we call \textit{the path length}. Note that the $L_2$ distance is not symmetric and the SMC and MCMC bounds depend on this quantity in opposite directions, and therefore  differ even for $S=1$ (importance sampling). 

For example when $\mu_0$ is heavy-tailed relative to $\pi$, ensuring $\|\pi/\mu_0\|_{L_2(\mu_0)}^2 \leq 1/\mathcal{E}$ and bounded SMC/IS error, $\|\mu_0/\pi\|_{L_2(\pi)}$ may be much larger than $1/\mathcal{E}$, slowing convergence of MCMC. However, the amount of computation required by SMC grows linearly in $S/\mathcal{E}$ whereas the bound for MCMC grows logarithmically in $\|\mu_0/\pi\|_{L_2(\pi)}$. This can be advantageous for MCMC when  finding a sequence of distributions that ensures $S/\mathcal{E}$ small is difficult.

The remainder of this section compares the  relative cost of moving from $\mu_0$ to $\pi$ for SMC versus MCMC. The first example investigates the problem of sampling from a spherical Gaussian target distribution, studying the path complexity with regard to the target precision, mean, and dimension. The second example considers the problem of sampling a general log-concave target distribution and uses an optimal path identified by~\cite{LovaszON3} to obtain an SMC path with low complexity. This bound achieves the same complexity as the best available results for MCMC.

\subsection{Gaussian example}\label{sec:gaussian}
Consider the problem of approximating expectations with respect to a $d$-dimensional spherical Gaussian target distribution $\pi(x) = N_d(\theta  1_d,\, I_d/\phi)$, where $\theta \geq 2$.  For $\phi\geq 1$, this problem is representative of many Bayesian inference problems with large sample sizes via the Bernstein-von Mises theorem.  A simpler version of this problem ($\theta=0$) was studied by~\cite{Marion2018}; however, results for the more challenging problem when $\theta\neq 0$ are now possible as Theorem~\ref{thm:total_bound} allows for unbounded importance %sampling 
weights.

We assume that the initial distribution for both SMC and MCMC is chosen to be standard Gaussian, $\mu_0 = N_d(0,\, I_d)$. The cost of MCMC, relative to the spectral gap, is given by
\begin{equation}
\label{eq:mcmc_gaussian}
    \bigO \big(\theta^2 d/(\phi(2-\phi))\big)
\end{equation}
assuming $\phi < 2$ and undefined for $\phi>2$ as the $L_2$ distance from $\pi$ to $\mu_0$ is unbounded (see Appendix).  This further highlights the difference in SMC and MCMC bounds discussed 
%in the previous section
above. For this problem, the MCMC bounds quickly become large when the starting distribution is flat relative to the target, whereas the SMC bounds are better when the starting distribution is more disperse than the target. We will consider two different choices of the interpolating distribution sequences for SMC which lead to bounds with improved complexity with respect to $\phi$, $\theta$ and $d$. These results are applicable for any $\phi\geq 1$.

A standard approach to constructing a sequence of interpolating distributions is a \textit{geometric mixture}, with $\mu_\beta(x) \propto \mu_0(x)^{1-\beta} \pi(x)^\beta$ for $\beta  \in [0,1]$. Such paths are commonly used to estimate ratios of normalizing constants, where they are sometimes referred to as \textit{power} or \textit{tempered} paths~\citep{GelmanMeng,Friel2014}. The path is specified by a sequence $\beta_0 = 0 \leq \beta_1 \leq \ldots \leq \beta_S=1$  controlling the rate at which the path moves from $\mu_0$ to $\pi$.  Choosing $\beta_s = \big(1+2/(\theta \sqrt{d})\big)^{s-1}\big/\big(\phi\cdot\theta\sqrt{d}\big)\wedge 1$ and $S=1+\big\lceil\frac{\theta \sqrt{d}}{2}\log\big(\phi^2\cdot\theta\sqrt{d}\big)\big\rceil$ ensures $\|\mu_s/\mu_{s-1}\|_{L_2(\mu_{s-1})}^2 \leq \bigO(1)$ and gives an upper bound on path length $S/\mathcal{E}$ of:
\begin{equation}\label{eq:tempered_gaussian}
  \bigO\big(\theta \sqrt{d} \cdot \log (\phi^2\cdot\theta \sqrt{d})\big)
\end{equation}
(see section~\ref{sec:appendix_gaussian}). This bound improves dimension dependence from $\bigO(d)$ to $\bigO(\sqrt{d}\log \sqrt{d})$ relative to the MCMC bound. We also see a  super-exponential improvement in dependence on the precision, from $\bigO\big(\frac{1}{\phi(2-\phi)}\big)$ to $\bigO(\log \phi)$, as well as an improvement in the location dependence from $\bigO(\theta^2)$ to $\bigO(\theta \log \theta)$. 

However an even better path exists, inspired by a result from~\cite{GelmanMeng}. Choose the first \linebreak $s_1=\ceil{3\sqrt{d}\log(\theta^2d)}$ distributions in the path to be $\mu_s=N_d\big(0, I_d/\phi_{1,s}\big)$ with $\phi_{1,s}=(1-1/\sqrt{9d})^s\vee \frac{1}{\theta^2d}$. The next distribution changes the location in a single step: $\mu_{s_1+1} = N_d(\theta 1_d , I_d \cdot \theta^2d)$. Finally, take the last $s_2 = \ceil{\sqrt{d}\log(d\theta^2\phi)}$ steps to be  $\mu_s = N_d(\theta 1_d, I_d/\phi_{2,s})$ with $\phi_2,s = \frac{1}{\theta^2d} \big(1+1/\sqrt{d}\big)^{s-s_1-1}\wedge \phi$. We call this the \textit{precision path}; % because 
it first decreases the precision in order to change locations in a single step. Since precision can be decreased exponentially quickly, this shortens the overall path, yielding an improved complexity in $\theta$ compared to varying the mean and precision simultaneously. More precisely, the precision path ensures $1/\mathcal{E}\leq 2$, giving a path length bound of (see section~\ref{sec:appendix_gaussian}):
\begin{equation}\label{eq:precision_gaussian}
  \bigO\big(\sqrt{d} \log(\phi\cdot\theta^2 d)\big)
\end{equation}
showing an improvement from $\bigO(\theta \log \theta)$ to $\bigO(\log \theta)$. This example highlights the potential speedup available using non-geometric paths, though in general finding such paths may be challenging.

\cite{GelmanMeng} derived an optimal path sampling estimator to estimate (log-) ratios of normalizing constants between normal distributions with different means. This optimal path also flattens the intermediate normal distributions by reducing their precisions, resulting in a similar improvement in complexity from $\mathcal{O}(\theta)$ (tempered path) to $\mathcal{O}(\log \theta)$. The similarity between good path-sampling and SMC paths arises due to the necessity of estimating intermediate ratios of normalizing constants to satisfy the one-step induction condition for SMC. In fact, when $d=1$ and $\phi=1$, a sufficiently fine discretization of the Gelman and Meng path yields the same complexity bound as the precision path. It is unlikely that this path is optimal for SMC, however, since the optimal path-sampling sequence from $\pi$ to $\mu_0$ is the reverse of the optimal path from $\pi$ to $\mu_0$, while this will not generally be true for SMC as the $L_2$ "distance" is asymmetric and optimal paths should reflect this asymmetry. 

\subsection{Log-concave target distributions}
Let $\pi(x)\propto q(x)$ be a log-concave target distribution on $\mathbb{R}^d$.  A function $q$ is 
%said to be 
\textit{strongly log-concave} if $q^{1-\alpha}(x)\cdot q^\alpha(y) < q\big(\alpha x + (1-\alpha) y\big)$ for $x, y \in \mathbb{R}^d$ and $\alpha\in(0,1)$. In general, bounds on the $\omega$-warm mixing times of Markov kernels targeting a 
%sequence of distributions obtained by tempering a log-concave distribution 
sequence of tempered log-concave distributions 
%obtained by tempering a log-concave distribution 
will have the same complexity at each step. For example, if we choose $K_s$ to be the Metropolis-adjusted Langevin algorithm (MALA)~\citep{roberts1996}, the complexity of the bound on the $\omega$-warm mixing time is independent of the temperature parameter~\citep{Dwivedi2018,Marion2018} and a similar result holds for other Markov kernels including the \textit{ball-walk} or \textit{hit-and-run walk} Markov kernels~\citep{LovaszGeometry}. Therefore, when the target distribution is log-concave, we can again focus on finding interpolating sequences that minimize the path length. 

Efficient path selection for tempered log-concave distributions has received substantial attention in the theoretical computer science literature
for estimating the volume of a convex body~\citep{LOVASZ_ON4,LovaszON3}.
%by sampling
%, where the volume of a convex body is estimated by sampling 
%from a sequence of tempered %distributions~\citep{LOVASZ_ON4,LovaszON3}. 
A key factor in volume computation is the $L_2$ distance between adjacent distributions, which controls the relative error when estimating the corresponding volume ratios. The following corollary follows from Theorem~\ref{thm:total_bound}, using the tempering path from~\cite{LovaszON3} and the bounds on the mixing time from~\cite{WuMALA}.

\setcounter{section}{2}
\setcounter{theorem}{1}
\begin{corollary}[SMC complexity for log-concave target distributions]\label{cor:log_concave}
Let $\pi(x)\propto q(x)$ be log-concave with mode $x^*$ and define $\kappa = L/m$ where for all $x,y\in\mathbb{R}^d$:
$$
-\frac{L}{2}\|x-y\|_2^2 \leq \log \frac{q(x)}{q(y)} - \nabla \log q(x)^T(x-y) \leq -\frac{m}{2}\|x-y\|_2^2
$$
Restrict $\pi$ to the ball $B$ of radius $ 4\sqrt{d/m}$ centered at $x^*$ and assume $\epsilon>2e^{-d}$. Choose $\mu_0 \propto \mathbb{1}_{B}(x)$ and $\mu_s(x) \propto \pi^{\beta_s}(x)\mathbb{1}_{B}(x)$ with $\beta_s = \frac{1}{d\kappa}\big(1+\frac{1}{\sqrt{d}}\big)^s$ for $s=1,\ldots,S=\ceil{\sqrt{d} \log(d\kappa)}$. Let $K_s$ be a MALA kernel with step size given in~\cite{WuMALA}. Then SMC provides a randomized approximation scheme in time %:
%
%$$\bigO^*(d\kappa).$$
$\bigO^*(d\kappa)$.
\end{corollary}
\setcounter{section}{3}
\setcounter{theorem}{1}

The notation $\bigO^*$ indicates the omission of logarithmic terms in $d$ and $\kappa$. The restriction to $B$ is used to bound the $L_2$ distance of the first step and has minimal impact on the results of our analysis as $\pi(B) \geq 1-\epsilon/2$; similar restrictions are common in the log-concave sampling literature. The assumption $\epsilon>2e^{-d}$ serves only to simplify the presentation.

The specified path ensures $\max_s \|\mu_s/\mu_{s-1}\|_{L_2(\mu_{s-1})} \leq \mathcal{E}^{-1}$ for $\mathcal{E} = e$ and therefore $S/\mathcal{E}=$ \linebreak$\bigO\big(\sqrt{d}\log(d\kappa)\big)$~\citep{LovaszON3}. Using the $\omega$-warm mixing time for the MALA kernel (Theorem 1 of~\cite{WuMALA}):
\begin{equation}
    \bigO\big(d^{1/2}\kappa \cdot \log^3\left(\max\{\kappa, d, \omega/\epsilon\}\right)\big)
\end{equation}
the result then follows from Theorem~\ref{thm:total_bound}.  Note %that 
this $\bigO^*(d^{1/2}\kappa)$ mixing time result of \cite{WuMALA} assumes a warm start; we can interpret the additional $\bigO^*(d^{1/2})$ complexity for the SMC path as the cost of starting from a feasible (explicit) initial distribution.  Indeed, this SMC bound is of the same complexity as recent results for MCMC with a MALA kernel starting from a feasible initialization~\citep{Dwivedi2018, Chen2020, Lee2020}, improving upon the previous best bound of $\bigO^* \big(d^{3/2} \kappa^2\big)$ for SMC~\citep{Marion2018}. (To the best of our knowledge, this is the fastest randomized approximation scheme for a log-concave target distribution.)  It is worth noting that the SMC algorithm uses a different target distribution at each step, allowing the MALA kernel - specifically the optimal %MALA 
step size - to vary, allowing
%. This allows  
for large MCMC steps initially when the target distribution is flat and smaller steps as the algorithm approaches the target distribution.  This is similar to the use of a time-inhomogenous %Markov 
chain to gradually refine the posterior distribution by~\cite{LOVASZ_ON4}.
%to the approach of ~\cite{LOVASZ_ON4}
%who use 
%a time-inhomogenous %Markov 
%chain to gradually refine the %posterior distribution.

\section{Adaptive path selection}
\label{sec:adaptive}
As noted in Section~\ref{sec:L2Bound},
selecting a path where a bound on the $L_2$ distance is known \textit{a priori}, let alone an optimal path, can be difficult in practice.  Adaptive SMC algorithms address this problem by using information from the current particle system to dynamically select the next distribution from a set of candidates.  
A common adaptive approach is to choose the next distribution by comparing the relative effective sample size (RESS) for different possible candidates~\citep{jasra2015levy, zhou2016toward, lin2013}.  
As noted in Section~\ref{sec:L2Bound},  the RESS can also be interpreted as a sampling estimate of the (inverse) $L_2$ distance. When this estimate is sufficiently accurate, it can be used to choose a path with bounded $L_2$ distance with high probability by ensuring ${E}_s>\mathcal{E}$ holds at each step with sufficiently high probability.
In this section, we describe a specific adaptive path SMC algorithm that uses %the 
RESS to dynamically choose the next distribution at each step. We begin with a discussion on the specification and selection of candidate distributions. %\scsout{We then extend the results of Theorem~\ref{thm:total_bound} to the adaptive setting, giving conditions under which the adaptive algorithm provably constitutes a randomized approximation scheme.} 

\subsection{Candidate distributions and path selection}\label{sec:path_assumptions}
A path is chosen from a family of distributions $\mu\left(x\mid \beta\right)$ indexed by a parameter $\beta \in \mathbb{B} \subset [0, 1]$; let $\V = \{\mu(x \mid \beta) : \beta \in \mathbb{B} \}$. To ensure the algorithm terminates, we will assume that every path starts with $\mu_0(x) = \mu\left(x \mid 0\right)$, ends with $\pi(x) = \mu\left( x \mid 1\right)$ and has finite length.  The problem of choosing a path is to select a set of distributions $\mu(x\mid \beta_1), \ldots, \mu(x\mid \beta_{S-1})$ so that the $L_2$ distance is controlled at each step of the algorithm.  For concreteness, we will illustrate ideas using the following running examples:

\begin{itemize}
\item[\textbf{Example 1:}] \textit{Let $\mu_0$ be an initial distribution and $\pi$ be a target distribution on $\X$ with respective unnormalized densities $q_{\mu_0}(x)$ and $q_\pi(x)$. Define the geometric mixture  $\mu(x\mid \beta) \propto q_{\mu_0}(x)^{1-\beta}  q_\pi(x)^{\beta}$ for $\beta \in [0,1]$. Then any finite sequence of $\beta$'s defines a single path.}
     \item[\textbf{Example 2:}] \textit{Let $\pi(x) \propto p(y_{1:K}|x) \pi_0(x)$ be a posterior distribution arising from a Bayesian model with likelihood $p(y_{1:K}|x) = \prod_{k=1}^K p(y_k|x)$ and prior distribution $\pi_0(x)$. Define the \emph{data-tempered} path as $\mu(x \mid \beta) \propto \prod_{k=1}^{\beta K} p(y_k|x)\pi_0(x)$ for $\beta \in \{0,1/K,\ldots, 1\}$.}
\end{itemize}

In both cases, we will assume that at each step, only increases in $\beta$ are considered.  In Example 1 the $L_2$ distance to $\pi$ provably decreases with increasing $\beta$ (see Lemma~\ref{Appendix E: monotone}).  In Example 2 the ordering is specified for computational ease and there is the potential for the distance to increase with certain increments of $\beta$.  In general, it is desirable to increase $\beta$ as quickly as possible - subject to the consideration that the $L_2$ distance is controlled - to minimize the overall number of SMC steps.

Associated with each $\mu \in \mathcal{V}$ will be a finite set of candidate distributions $\mathcal{V}\left(\mu\right) \subset \mathcal{V}$ and at each step of the algorithm the next target distribution will be chosen from this set of candidates.  Suppose that the algorithm is currently at step $s$ with candidate distributions $\V\left(\mu_s\right)$.

\begin{itemize}
    \item[\textbf{Example 1:}] \textit{For $\beta_s$, the parameter value at step $s$ of the algorithm, define the candidate distributions for step $s+1$ by $\V\left(\mu \left(x \mid \beta_s \right)\right) = \left\{\mu \left(x \mid  \beta_s + \frac{m}{M}(1-\beta_s)\right) : m\in \{1,\ldots,M\}\right\}$.}
     \item[\textbf{Example 2:}] \textit{For $\beta_s=\frac{m_s}{K}$, the current fraction of data included at step $s$ of the algorithm, define the candidate distributions by $\V\left(\mu \left(x \mid \beta_s \right)\right) = \left\{\mu \left(x \mid\beta_s + m/K\right) : m \in \{1,\ldots,K-m_s\} \right\}$, where $K-m_s$ is the number of additional data points to be incorporated.}
\end{itemize}

We denote the set of candidate distributions from $\mu_s$ at step $s$ by $\V\left(\mu_s\right)=\{\nu_{s,m}\}_{m=1}^{M}$. %with the subscript $s$ indicating that $\nu_{s,m}$ is a candidate distribution for step $s+1$ and that the set of candidates depends on $\mu_s$.  
Each candidate $\nu_{s,m}$ has corresponding importance weight function $w_{s,m}(x) \propto  \nu_{s,m}(x) / \mu_s(x)$. 

\begin{itemize}
    \item[\textbf{Example 1:}] \textit{For the geometric mixture the importance weight functions are\linebreak $w_{s,m}\left(x\right) = \left(q_\pi(x)  /q_{\mu_0}(x)\right)^{\frac{m}{M}\left(1-\beta_s\right)}$.}
     \item[\textbf{Example 2:}] \textit{For the data-tempered path the importance weight functions are the additional likelihood terms $w_{s,m}(x) = \prod_{k=1}^{m} p(y_{\beta_s K + k}|x)$.}
\end{itemize}

Any adaptive-path SMC algorithm requires a rule for selecting the next step $\mu_{s+1}$ of the path from the candidate set $\V(\mu_s)$ at each step $s$.  Denote this selection rule by the function $r:\X^N \times \V \rightarrow \V$. Evaluated at a specific distribution $\mu\in\V$, this function maps the sampled particles to one of the distribution's candidates: $r\left(\,\cdot \, ,\; \mu \right): \X^N \rightarrow \V\left(\mu\right)$. We consider selection rules guided by the RESS ${E}_{s,m}$ of each candidate step:
\begin{equation}
%    \begin{split} &
{E}_{s,m} = W_{s,m}^2 / {V}_{s,m} %\\ &
\qquad \quad 
W_{s,m} = N^{-1}\sum_{i=1}^N w_{s,m}({X}_s^n) %\\ &
\qquad \quad 
{V}_{s,m} = N^{-1}\sum_{i=1}^N w_{s,m}^2({X}_s^n). %\\
%  \end{split}
\label{Eqn:ESSDecomposition}
\end{equation}
Specifically, we consider the following step selection rule:
\begin{enumerate}
\item 
Set $\mu_{s+1} = r\big( X_s^{1:N}, \mu_s\big) = \nu_{s, m^*}$ where
\begin{equation}\label{eq:selection}
%    \begin{split}
%        \text{Set} \; \mu_{s+1} &= r\left( X_s^{1:N}, \mu_s\right) = \nu_{s, m^*} \text{ where }\\
        %
m^* %& 
= \max\left\{ m \in \left\{1,\ldots,M\right\}: \big({E}_{s,m}\geq \mathcal{E}\big)\cap \big({V}_{s,m} \geq \mathcal{C}\big) \text{ or } m=1\right\}
%    \end{split}
\end{equation}
\end{enumerate}
To specify the algorithm the n, the user specifies the two quantities
%which control the behavior of the rule, 
$\mathcal{E}$ and $\mathcal{C}$. The quantity $\mathcal{E}\in(0, 1)$ is a user-specified lower bound on the RESS; $1/\mathcal{E}$ is interpreted as a target $L_2$ distance step-size. Large values of $\mathcal{E}$ will yield shorter paths (longer steps), but increased $\Var{(\hat{f})}$, whereas small values will lead to longer paths (shorter steps) but lower $\Var{(\hat{f}})$. Step selection based on the RESS is commonly used as a heuristic for automatically selecting a path~\citep{jasra2015levy, zhou2016toward, lin2013}.

The additional requirement that ${V}_{s,m}\geq \mathcal{C}$ for pre-specified $\mathcal{C}$ is introduced to ensure that $1/{E}_{s,m}$ approximates $\|\nu_{s,m}/\mu_s\|_{L_2(\mu_s)}^2$ with bounded relative error. Small values of $\mathcal{C}$ may allow for bigger steps; however, this advantage must be balanced against the number of particles required to achieve this relative error bound. We show in Section~\ref{sec:adaptive_algorithm} that the computational complexity of the algorithm grows as $\bigO \big(\mathcal{C}^{-2}\big)$. An alternative step selection criteria omitting this `$\mathcal{C}$' condition is given in the appendix (Section~\ref{Sec:AltThm2}).

If no candidate distribution meets these requirements the algorithm defaults to $\mu_s=\nu_{1, s}$. For our bounds to hold, we therefore require that $\nu_{1,s}$ satisfy the conditions $\Lp{\nu_{1,s}}{\mu_s}{2} \leq 2 \mathcal{E}^{-1}$. This requires the ability to choose an increment $\Delta \beta$ which is ``sufficiently small".  For 
Example 1, 
Lov\'{a}sz and Vempala~\cite{LovaszON3} show how an explicit upper bound on the $L_2$ distance can be obtained for a special case of the geometric mixture when $\pi$ is a log-concave distribution and $\mu_0$ is a tempered version of $\pi$.  More generally, even when  $q_{\mu_0}(x)$ and $q_{\pi}(x)$ are not log-concave, the $L_2$ distance decreases monotonically in $\frac{1}{\Delta \beta}$ %$\Delta \beta^{-1}$ 
for geometric paths (Lemma~\ref{Appendix E: monotone}). In this case, although the rate at which the $L_2$ distance grows with $\Delta \beta$ depends on $q_{\mu_0}(x)$ and $q_{\pi}(x)$, the monotonicity guarantees that a sufficiently small $\Delta \beta$ exists, so 
%when 
if the condition $\Lp{\nu_{1,s}}{\mu_s}{2} > 2 \mathcal{E}^{-1}$ fails the algorithm can simply be restarted with an $M^* > M$.
(Note %that 
the bounds in Theorem~\ref{thm:adaptive_total_bound} grow %at most logarithmically in $M$.
as $\log(M)$.) %On the other hand
However, %for 
the data tempering paths of Example 2 %there is no %such 
have no 
guarantee of monotonicity.  In Section~\ref{sec:linear_regression} we provide a modification of the data tempering approach which addresses this weakness. In either case, we assume that a sufficiently small $\Delta \beta$ exists in the candidate set at the start of the algorithm.  Online refinement of $\Delta \beta$ when this condition is violated would constitute a minor modification of the algorithm; however, our proof currently does not extend to this situation.

\subsection{Adaptive Path Selection SMC}
\label{sec:adaptive_algorithm}
We now state the adaptive path SMC algorithm.  As before the adaptive path SMC algorithm is initialized by drawing $N_0$ independent samples $X_0^{1:N_0}$ from $\mu_0$. The realizations of these particles are denoted by $x_0^{1:N} = \big( x_0^0,...,x_0^{N_0} \big)$. Step $s$ of the algorithm proceeds as follows:
\begin{itemize}
\item[(i)] Select a candidate distribution $\mu_s \in \V\left(\mu_{s-1}\right)$ by setting $\mu_s= r\big(x_{s-1}^{1:N}, \mu_{s-1}\big)$.
\item[(ii)]  Assign each particle 
%an importance 
weight equal to the unnormalized density ratio:
%
%    $$w_s\big(x_{s-1}^{n}\big) = \frac{q_s\big(x_{s-1}^{n}\big)}{q_{s-1}\big(x_{s-1}^{n}\big)}.$$
$$w_s(x_{s-1}^n) = q_s(x_{s-1}^n)/q_{s-1}(x_{s-1}^n).$$
\item[(iii)] Sample a new set of particles $\tilde{X}_s^{1:N_s}$ with replacement according to the weights: %(multinomial resampling):
$$ \Pr\big(\tilde{X}_s^n = x \;\big|\; X_{s-1}^{1:N_s} = x_{s-1}^{1:N_s}\big) \propto \sum_{n=1}^{N_s} w_s(x_{s-1}^n) \cdot \delta_{x_{s-1}^n}(x).$$
\item[(iv)] Apply $t_s$ steps of the kernel $K_{\mu_s}$ to each re-sampled particle, producing $X_s^{1:N_s}$: 
$$ \Pr\left(X_s^n \in dx \;\mid\; \tilde{X}_s^n =\tilde x_s^n\right) = K_{\mu_s}^t\big(\tilde x_s^n, \;dx \;\big).$$
\end{itemize}
Steps (\textrm{i}) - (\textrm{iv}) are repeated until $\pi$ is selected as the candidate distribution, which must happen eventually since each path is assumed to end with $\pi$ and has finite length. The step number $S$ at which this occurs is random; we denote the penultimate step as $S-1$. At the final step $S$ of the algorithm steps (\textrm{i}) - (\textrm{iv}) are repeated producing a set of particles $X_S^{1:N_S}$ which are approximately distributed according to the target distribution $\pi$. The SMC estimate of $\pi(f)$ is $\hat{f} = \frac{1}{N_{S}} \sum_{n=1}^{N_S} f\big(x_{S}^{n}\big)$.

\subsection{Discussion of Theorem~\ref{thm:adaptive_total_bound}}
Theorem~\ref{thm:adaptive_total_bound} establishes that the adaptive algorithm described in Section~\ref{sec:adaptive_algorithm} provides a randomized approximation scheme for $\pi f$ under the conditions \asref{as:3}-\asref{as:5}.

\asref{as:3} is needed to provide conditions under which the $L_2$ distance can be accurately estimated. \asref{as:4}, discussed at length at the end of Section~\ref{sec:path_assumptions}, ensures that the algorithm does not terminate prematurely due to a lack of viable candidate distributions.  \asref{as:5} restates key properties of the paths (from Section~\ref{sec:path_assumptions}) to make all conditions explicit; these properties generally hold in practical applications of SMC.  The requirement in \asref{as:5} that the number of candidates be bounded is necessary to control the error of each ${E}_{s,m}$, which requires coupling and concentration inequalities for each possible candidate. This requirement is often satisfied in practice, even when the number of candidates is theoretically larger.  For example, while the geometric mixture or tempering are often described as using a continuous temperature ladder, in practice adaptive algorithms select candidates by searching over a finite number of candidates. 

As noted in Section~\ref{sec:L2Bound}, the proof of Theorem~\ref{thm:adaptive_total_bound} is modifies the argument of Theorem~\ref{thm:total_bound} to address the randomness introduced by the adaptive nature of the algorithm.  At each step, the selected candidate distribution is random since it is chosen based on the sampled particles. This presents a challenge, as Corollary~\ref{cor:iteration} requires $\|\mu_s/\mu_{s-1}\|_{L_2(\mu_{s-1})}^2 < \mathcal{E}^{-1}$ to hold for all steps $s$, regardless of which candidate is chosen. Instead, the conditions imposed by our selection rule ensure that the selected step satisfies this bound on the $L_2$ distance  with high probability. The randomness in the choice of interpolating distributions means that the total number of steps $S$ in the selected path is also random.  We show that, by gradually increasing the size of the particle system $N_s$ and the number of Markov transitions $t_s$ at each step as specified in the algorithm, a bound similar to that in Theorem~\ref{thm:total_bound} can be obtained for any finite $S$ at which the algorithm terminates.  

The full proof of Theorem~\ref{thm:adaptive_total_bound} is given in Appendix~\ref{sec:thm2_proof}. 

\section{Empirical results}\label{sec:empirical}

We investigate the empirical performance of the adaptive step-size SMC algorithm on two non-trivial target distributions where the $L_2$ distance can be evaluated exactly. Our goals are twofold. First, choosing $N_s$ and $t_s$ to meet the requirements of Theorem~\ref{thm:adaptive_total_bound} is generally difficult due to the challenge of bounding the mixing time. Therefore, it is sensible to verify whether a naive implementation of the algorithm might maintain controlled $L_2$ distances at each step. Second, Theorem~\ref{thm:adaptive_total_bound} does not address the relative optimality of the adaptively chosen path, guaranteeing accurate estimation but not optimal path length. This empirical study allows us to compare adaptively chosen paths to an ideal path which maintains a fixed step size of $\mathcal{E}$. Finally, we provide an example where data tempering may lead to candidate sets where the conditions in~\eqref{eq:selection} cannot be satisfied. To address this problem, we introduce a hybrid path construction method that ensures that the conditions in~\eqref{eq:selection} can be met at each stage of the algorithm. Empirically, we find that this hybrid approach decreases the overall path length and leads to paths of near-optimal length for a given step size. For these reasons, we recommend that the hybrid approach be preferred over the data tempering approach in practice.

\subsection{Example: Ising model}
Consider the well-known  mean field Ising model, originally developed as a model of ferromagnetism in statistical physics. The $D$-dimensional model takes values 
$(x_1,\ldots,x_D) \in \X = \{-1,1\}^{D}$ for binary ``spins'' $x_d$ with probability
\begin{equation}
    \pi(x|\alpha) \propto \exp \Big(\frac{\alpha}{2D}\big(\sum_{d=1}^D x_d\big)^2\Big).
\end{equation}
When $\alpha > 0$, the high probability  configurations are those where the spins are mostly the same. The hyperparameter $\alpha$ controls the strength of this effect.  Related models have been used in machine learning for image processing~\citep{Salakhutdinov08learningand} and in Bayesian statistics for modeling spatial dependence~\citep{Bannerjee2015, Besag1974}.

Sampling from the Ising model has received considerable attention~\citep{binder2002, Propp1996, Jerrum1993}.  A key characteristic of the model is the phase transition in $\pi$ as $\alpha$ approaches critical temperature $\alpha_0$, exhibited in the distribution of the magnetization  $M=\sum_{d=1}^D x_d$, which rapidly changes from concentrated about $0$ to dispersed to the extremes near $M=D$ and $M=-D$.  This rapid change in behaviour makes sampling from the Ising model challenging when $\alpha>\alpha_0$; for example it is difficult for random-walk MCMC methods such as Glauber dynamics to move between the two modes. Tempering approaches have proven to be a solution for sampling from this distribution (%see~
\cite{Woodard09rapid} and references therein). The selection of an appropriate temperature ladder is crucial to the success of tempering, and subsequently selecting a temperature ladder has received substantial attention in the tempering literature~\citep{Predescu2004, Rathore2005, Stefankovic2007, Vousden2015}. Here we demonstrate empirically that temperature selection using an adaptive step-size SMC approach achieves nearly optimal performance in the sense of minimizing the number of steps (temperatures) for a given $\mathcal{E}$.

For $D\in\{10,50,250\}$ and $\alpha=2$ we performed SMC using the geometric mixture given in Section~\ref{sec:path_assumptions} with $\mu_0$ uniform on $\X$. Markov transitions are made according to the Glauber dynamics (Gibbs sampling), scanning through each site in a randomly chosen order and drawing a new spin from its conditional distribution. We set $\mathcal{E}=0.5$ and used $N=1,000$ particles for each simulation. This SMC procedure was repeated $1,000$ times to assess variability. For the chosen values of $D$, the the marginal distribution of the magnetization can be computed directly and the corresponding $L_2$ distance between marginal distributions can be evaluated numerically. 

\begin{figure}%[H]
\centering
\includegraphics[width=0.95\textwidth]{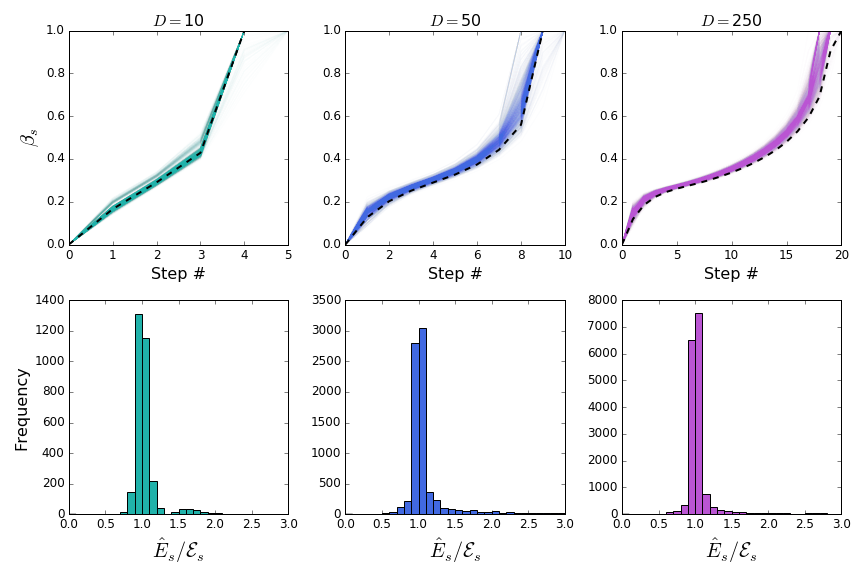}
\captionsetup{width=.90\linewidth}
\caption{Results of the empirical path selection for the mean field Ising model. The top row shows the distribution of paths chosen by the adaptive approach, with the optimal path given by the dotted black line. The bottom row shows the distribution of the estimated $L_2$ distance relative to the true distance.}
\label{fig:ising} 
\end{figure}

Figure~\ref{fig:ising} shows a comparison of the relative error of the SMC estimate $\mathcal{E}^{-1}$ and the true $L_2$ distance. We also compared the adaptively chosen path to the temperature ladder satisfying exactly $\|\mu_s/\mu_{s-1}\|_{L_2(\mu_{s-1})}^2 = 1/\mathcal{E}$ at each step. In general, the adaptively chosen paths follow closely the optimal path, being of comparable length and displaying similar curvature near the critical temperature, where they take small steps as the target distribution is changing rapidly. Estimated values of $\|\mu_s/\mu_{s-1}\|_{L_2(\mu_{s-1})}^2$ are generally quite accurate. More importantly, the induction condition prescribed by Lemma~\ref{lemma:l2_selection} that requires $\Lp{\mu_{s+1}}{\mu_s}{2} \leq 2\mathcal{E}^{-1}$ is achieved at every step of the algorithm for every simulation. The selection criteria in~\eqref{eq:selection} achieves good path selection for this problem, suggesting that the modified criteria \eqref{eq:selection} may perhaps not be required in this case.

\subsection{Bayesian linear regression}\label{sec:linear_regression}
Our second example demonstrates the behaviour of adaptive SMC using data tempering. Consider a Bayesian linear regression model with $Y = X\beta + \epsilon$, where $Y\in \mathbb{R}^K$ is an observed response vector, $X\in \mathbb{R}^{K\times D}$ a matrix of covariates, $\beta\in \mathbb{R}^D$ an unknown coefficient vector and $\epsilon \sim N(0, I_K\cdot\sigma^2)$ a vector of observation noise. We fit this model to the 'white wines' data set from the UCI machine learning repository~\citep{Dua:2017}, which consists of $M=4898$ observations of wine quality and $D=11$ physicochemical predictors. Before analysis, the data was centered and scaled. We adopt a normal inverse-gamma prior with $\pi_0(\beta\mid\sigma^2) \propto N\big(0, \sigma^2\big(X^T X)^{-1}/K\big)$ and $\pi_0(\sigma^2)\sim \text{Inv-Gamma}(4,4)$. This prior is conjugate, allowing analytic calculation of the $L_2$ distance between SMC steps for comparison (see Section~\ref{sec:regression_theory}).

Simulation from the posterior distribution of this model was performed using the data tempering approach described in Section~\ref{sec:path_assumptions}. During the initial phase of the algorithm, the target distribution changes rapidly as observations are added, making it difficult to obtain transitions with sufficiently high relative effective sample size. As a result, instead of choosing $\mu_0=\pi_0$, we let $\mu_0 = \pi_{200}$ for $\pi_n \propto p(Y_{1:n} \mid \beta, \sigma^2, X_{1:n}) \cdot \pi_0(\beta, \sigma^2)$.  This starting point can be easily obtained in practice, either by MCMC or via a geometric path from the prior %distribution, 
(or here, %, in this case, by 
direct sampling due to conjugacy). %The Markov kernels are chosen %to be 
as Gibbs samplers, alternating draws of $\beta\mid\sigma^2$ and $\sigma^2\mid\beta$. $1,000$ SMC runs each using $N=1,000$  particles were conducted, each adaptively choosing a path using $\mathcal{E}=1/2$. Observation ordering was permuted randomly between each trial to assess the sensitivity of the procedure to the dataset ordering. 

\begin{figure}%[H]
\centering
\includegraphics[width=0.95\textwidth]{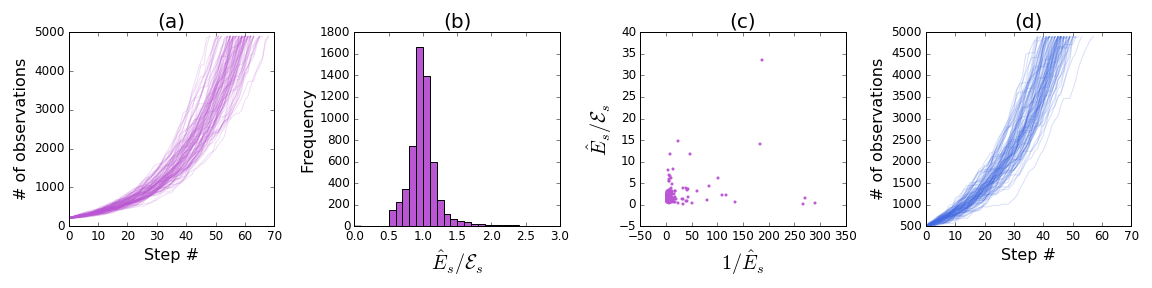}
\captionsetup{width=.9\linewidth}
\caption{Adaptive path selection for the linear regression model. (a) Distribution of paths chosen by the adaptive approach. (b) Distribution of the estimated $L_2$ distance relative to the true $L_2$ distance for transitions with $1/{E}_s\leq 2$. (c) Relative size for steps with bad transitions, i.e. $1/{E}_s\geq 2$. (d) Distribution of paths chosen by the hybrid algorithm} 
\label{fig:regression} 
\end{figure} 

 The results of the simulation experiment are %displayed 
 shown in Figure~\ref{fig:regression}. The number of steps required by the adaptive %procedure 
 algorithm 
 grows logarithmically %with 
 in the number of observations, 
 %providing evidence of 
 indicating 
 the efficiency of the data tempering approach. But
 %However 
 this advantage comes at a cost; a key difference between tempering and data tempering is that %the 
 data tempering 
 %approach 
 is more likely to fail at controlling the $L_2$ distance. This occurs when the next observation in the data tempering sequence results in a transition with ${E}_{s,1} \leq \mathcal{E}$ leading to uncontrolled error.  Across all experiments, nearly $5\%$ of the steps resulted in no candidate distributions satisfying ${E}_{s,1} \geq \mathcal{E}$; this tends to occur when moving to a high-leverage point, %which results 
 resulting in large changes to the posterior. The relative error of these steps is shown in Figure~\ref{fig:regression}(c). These steps are characterized by large $L_2$ distances, which are often significantly underestimated by the RESS.

This problem occurs because sequential introduction of data points has effectively established a minimum SMC step size that is too large, leading to an insufficiently rich set of possible paths. When the algorithm is forced to take a large jump for lack of alternatives, it may take smaller steps at subsequent iterations. This can allow the sampler to recover from a large step at the cost of additional computation. To address this problem, we introduce here a hybrid path that combines the computational advantages of the data tempering approach with the rich set of paths afforded by tempering. This hybrid path generally ensures that a satisfactory transition can be made at each step of the algorithm and we observe in Figure~\ref{fig:regression} that this leads to shorter, more efficient path-lengths on average. 

\textbf{Hybrid path for sequential data:} Assume the same setting as the data tempered path and suppose $\mu_{s-1}(x) \propto p(y_{1:k_{s-1}}|x)\pi_0(x)$. First, consider the move to $\nu_{s,1}(x) \propto p(y_{1:k_{s-1}+1}|x) \cdot \pi_0(x)$. If ${E}_{s,1} \geq \mathcal{E}$, consider additional candidate distributions using the data tempering approach. If not, choose candidate distributions in the same manner as the geometric path, selecting from the family $\eta \propto p(y_{1:k_{s-1}} \mid x)p(y_k \mid x)^{\beta} \pi_0(x)$ for $\beta \in[0,1]$. Several tempering steps may be required to reach $p(y_{1:k_s+1} \mid x)$, at which time we again consider both data-tempering and tempering moves.

The hybrid path provides a solution to the problem of unacceptably large transitions induced by influential data points. The tempering steps allow for smoother interpolation from the prior to the posterior by effecting smaller changes in the posterior, effectively allowing the addition of ``fractional" data points when refining the step size. 

The paths obtained by application of this hybrid approach to the Bayesian linear regression example are shown in Figure~\ref{fig:regression}(d). The flexibility to introduce \textit{fractional} data points results in the algorithm being able to satisfy the adaptive criteria at each step, with no failed transitions and correspondingly accurate estimation of the $L_2$ distances (not shown). As described above, the paths obtained tend to be shorter than those obtained from the traditional data tempering approach. This arises from cascading effect, since a poor transition leads to increased error in the estimation of the $L_2$ distance, often resulting in unnecessarily conservative future transitions.

\section{Conclusion}
Theorems~\ref{thm:total_bound}~and~\ref{thm:adaptive_total_bound} provide finite-sample guarantees on the performance of SMC.   Sections~\ref{sec:path_examples} and~\ref{sec:empirical} show that these results can be used to analyze and inform the selection of distribution paths, and demonstrate the importance of path selection on the algorithm's performance. Approximately optimal selection of paths can dramatically improve the efficiency of SMC estimators; for the Gaussian and log-concave examples in Section~\ref{sec:path_examples} the resulting bounds match the best available bounds for MCMC. When an efficient path is not known prior to running the algorithm, our results show that paths with bounded error guarantees can be obtained during sampling by adaptively choosing steps using the RESS to estimate the $L_2$ distance. Although we currently have no theoretical guarantees of optimality, empirical results on the examples presented in Section~\ref{sec:empirical} suggest that this approach can lead to near-optimal paths. Further approaches to automatic construction of efficient paths, such as those in~\cite{yao2020adaptive} or~\cite{marionThesis}, may provide further improvements for practical applications.

The theorems presented in this paper require weaker conditions than those used in previous finite sample results for SMC~\citep{Marion2018}, which assumed bounded weights and a lower bound on the 
%ratio of normalizing constants
normalizing constant ratio.  Those assumptions bound the $L_2$ distance as $\Lp{\mu_s}{\mu_{s-1}}{2} \leq \sup_{x\in\X} w_s(x)^2 \cdot z_s^2/z_{s-1}^2 $, satisfying~\asref{as:1} for Theorem~\ref{thm:total_bound} and~\asref{as:3} and~\asref{as:4} of Theorem~\ref{thm:adaptive_total_bound}. These weaker assumptions enable us to address a broader class of problems as demonstrated  in Section~\ref{sec:path_examples}. Theorem~\ref{thm:total_bound} of the current paper applies more broadly than Theorem~\ref{thm:adaptive_total_bound}, but the result for the adaptive algorithm requires the additional assumption of  bounded weights. The advantage of Theorem~\ref{thm:adaptive_total_bound} over Theorem~\ref{thm:total_bound} is that when the bound on the $L_2$ distance required by both is not tight, the adaptive algorithm may %be able to 
take larger steps which more closely approximate the target step size.

\fund
JM was partially supported by the National Science Foundation (NSF) grant DMS-1638521 to the Statistical and Applied Mathematical Sciences Institute (SAMSI) SAMSI and by NSF research traineeship grant DMS-1045153. SCS was supported by NSF grant DMS-1638521 to SAMSI and by NSF grant DMS-1407622.

\bibliographystyle{APT}
\bibliography{mybib.bib}

%%%%%%%%%%%%%%%%%%%%%%%%%%%%%%%%%%%%%%%%%%%%%%
%% Example with single Appendix:            %%
%%%%%%%%%%%%%%%%%%%%%%%%%%%%%%%%%%%%%%%%%%%%%%

\appendix
\section{Proof of Theorems}
\subsection{Proof of Theorem~\ref{thm:total_bound}}\label{sec:thm1}
Our proof follows closely the coupling argument of~\cite{Marion2018}, by constructing a set of independent random variables $\bar{X}_s^{1:N}$ for $s=1,\ldots,S$ satisfying $\bar{X}_s^n \sim \mu_s$ with $\mu_S = \pi$\, such that $\Pr(X_S^n = \bar{X}_S^n) \geq 1-\delta$. The coupling construction proceeds by showing that at each of the algorithm step $s=0,\ldots,S$ the following events occurs with high probability:
\begin{equation}\label{eq:thm_conditions}
%\begin{split}
\A{s} %& 
= \big{\{}X_s^{1:N} = \bar{X}_s^{1:N}\big{\}} %\\
\qquad \quad
\B{s} %& 
= \big{\{}  W_{s+1} > \frac{2}{3}\mu_s\left(w_{s+1}\right) \big{\}} %\\
\qquad \quad 
\C{s} %& 
= \A{s} \cap \B{s},
%\end{split}
\end{equation}
%
%\scsfootnotetext{I don't think $W_s$ has been defined yet; definition for $\bar{W}_s$ appears in paragraph after Eq~\eqref{Eqn:PrCs} below, but that's for coupled particles.  I don't see a definition for $W_s$.     Eq~\eqref{Eqn:ESSDecomposition} seems closest, but seems like it should really be defined around Eq~\eqref{eq:selection1} or in Section~\ref{Sec:SMCIntro}. }
where $W_{s+1} := \frac{1}{N}\sum_{i=1}^N w_{s+1}(X_s^n)$ and 
$\A{s}$ is the event that the SMC particles at step $s$ couple to a set of independent random variables $\bar{X}_s^{1:N}$ satisfying $\bar{X}_s^n \sim \mu_s$. $\B{s}$ is the event that the empirical average of the weights do not underestimate their expectation by too much. $\C{s}$ is the event that both occur.  Lemma 5 and Corollary 5.1 from~\cite{Marion2018} combine to prove the following lower bound under the assumptions $w_s(x) \leq W$ and $z_s/z_{s-1} \leq Z$ for all $s$:
\begin{equation}
\label{Eqn:PrCs}
    \Pr \left( \C{s} \right) \geq \left(1-\delta\right)\cdot \Pr\left (\C{s-1} \right)- \delta'
\end{equation}
where $\delta=1/8S$ and $\delta'=1/64S$, for $S$ the number of SMC steps.  The proof in~\cite{Marion2018} uses H\"oeffding's inequality to establish concentration of the average weight of the coupled particles, $\bar{W}_{s+1} = \frac{1}{N}\sum_{i=1}^N w_{s+1}(\bar{X}_s^n)$ around it's expectation, ensuring %that 
with high probability:
$$ |W_{s+1} - \mu_s(w_{s+1}) | \leq \mu_s(w_{s+1})/3$$
The proof of Theorem~\ref{thm:total_bound} in the current paper follows by substituting a one-sided Bernstein inequality (a lower bound suffices) in place of H\"oeffding's inequality.

\begin{lemma}\label{lemma:concentration}
Let $\bar{X}_s^{1:N}$ be independent random variables each with distribution $\mu_s$, and let \linebreak $\bar{W}_{s+1} = \frac{1}{N}\sum_{i=1}^N w_{s+1}(\bar{X}_s^n)$. Assume  $\|\mu_s/\mu_{s-1}\|_{L_2(\mu_{s-1})}^2 \leq \mathcal{E}^{-1}$, then for any $N\geq 18\log\left(1/\delta'\right)/\mathcal{E}$ we have
\begin{equation}
\Pr\Big(  \bar{W}_{s+1} \leq \frac{2}{3}\mu_s\left(w_{s+1}\right) \Big) < \delta'.
\end{equation}
\end{lemma}

\begin{proof}
Let $Z^{1},\ldots,Z^N$ be independent and identically distributed random variables such that $Z^i \leq b$. The one-sided Bernstein inequality gives
\begin{equation}
    \Pr\Big(\frac{1}{N}\sum^N_{i=1} Z_i - \E[Z_i] \geq t \Big) \leq \exp\Big\{-\frac{N t^2}{2\big(\E[Z^2_i] + \frac{b t}{3} \big)} \Big\}.
\end{equation}
Note that $ \|\mu_s/\mu_{s-1}\|_{L_2(\mu_{s-1})}^2= \frac{\mu_{s-1}(w^2_s)}{\mu_{s-1}(w_s)^2}$. Applying Bernstein's inequality to $-w_s(\bar{X}^{n}_{s-1}) < 0$, we get %obtain
\begin{equation}
%    \begin{split}
\Pr\Big(\frac{1}{N}\sum^N_{n=1} w_s(X^{n}_{s-1}) \leq \mu_{s-1}(w_s) \cdot \frac{2}{3} \Big) %&
\leq \exp\Big\{-\frac{N}{18  \|\mu_s/\mu_{s-1}\|_{L_2(\mu_{s-1})}^2 } \Big\} %\\
        %
        %&
\leq \exp\Big\{-\frac{N \mathcal{E}}{18 } \Big\}. 
%    \end{split}
\end{equation}
Setting $N > 18 \log(\frac{1}{\delta'}) / \mathcal{E}$ gives the result.
\end{proof}

Substituting Lemma~\ref{lemma:concentration} above into the proof of Lemma 5 of~\cite{Marion2018}, with the rest of the proof remaining the same, allows us to obtain the following corollary. 

\begin{corollary}
\label{cor:iteration}
Suppose $\Pr\left(\C{s-1} \right) \geq \frac{3}{2 \omega}$ for some $\omega > \frac{3}{2}$.  Fix $0 < \delta < 1$ and $0 < \delta' < 1$ and assume  $\|\mu_s/\mu_{s-1}\|_{L_2(\mu_{s-1})}^2 \leq \mathcal{E}^{-1}$. Then for any $N\geq - \mathcal{E}\log(\delta')/18$ and $t \geq \tau_s\big(\frac{\delta}{N}, \omega\big)$ we have
\begin{equation}
    \Pr \left( \C{s} \right) \geq \left(1-\delta\right)\cdot \Pr\left (\C{s-1} \right)- \delta'.
\end{equation}
\end{corollary}

The proof of Theorem~\ref{thm:total_bound} then follows immediately by the argument in~\cite{Marion2018} with Corollary 5.1 there replaced by Corollary~\ref{cor:iteration} above.  We note that other concentration inequalities for independent random variables could be used to show additional variants of Theorem~\ref{thm:total_bound}.  For example, results similar to~\cite{Chatterjee2018} could be used to prove a bound using the Kullback-Leibler divergence instead of the $L_2$ distance.

\subsection{Proof of Theorem ~\ref{thm:adaptive_total_bound}}\label{sec:thm2_proof}
As in the proof of Theorem~\ref{thm:total_bound}, we show that at each step $s=0,\ldots,S$ of the algorithm the following events occurs with high probability:
\begin{equation}
%\begin{split}
\A{s} %& 
= \big{\{}X_s^{1:N_s} = \bar{X}_s^{1:N_s}\big{\}} %\\
\qquad \quad 
\B{s} %& 
= \big{\{} W_{s+1} > \frac{2}{3}\mu_s\left(w_{s+1}\right) \big{\}} %\\
\qquad \quad 
\C{s} %& 
= \A{s} \cap \B{s}.
%\end{split}
\end{equation}
These conditions are the same as those required in Appendix~\ref{sec:thm1}, with a slight change allowing for the number of particles $N_s$ to depend on the step $s$.  As in the proof of Theorem~\ref{thm:total_bound}, we show that following inductive conditions hold at each step of the algorithm:
\begin{itemize}
\item[Condition 1:] $\Pr\left(\A{s}\right) \geq (1-\delta_s) \cdot \Pr\left(\C{s-1}\right).$
\item[Condition 2:] $\Pr \left(\C{s} \right) \geq \left(1-\delta_s\right)\cdot \Pr\left (\C{s-1} \right)- \delta_s'$.
\end{itemize}
The terms $\delta_s,\delta_s' \in (0, 1)$ are error bounds at each step that are related to the number of samples $N_s$ and Markov kernel transitions $t_s$, respectively. Compared with the proof of Theorem~\ref{thm:total_bound}, the dependence of $\delta_s$ and $\delta_s'$ on $s$ here accommodates the unknown (random) number of steps $S$ in the adaptive algorithm. Condition 1 follows from Lemma 2 in~\cite{Marion2018}. We re-state that result here in the form of the following lemma to make the dependence between $\delta_s$ and $N_s$ clear.

\begin{lemma}\label{lemma:coupling2}
 Suppose $\Pr\left(\C{s-1} \right) \geq 3/4$. Define $\tau\left(\;\cdot\;, 2\right) = \sup_{\mu \in \V} \tau_\mu\left(\;\cdot\;, 2\right)$. Then for any $0<\delta_s<1$ and $t_s \geq \tau\big(\frac{\delta_s}{N_s}, 2\big)$:
\begin{equation}
    \Pr\left(\A{s}\right) \geq (1-\delta_s) \cdot \Pr\left(\C{s-1}\right) .
\end{equation}
\end{lemma}
The proof of Condition 2, the inductive condition, is complicated by the fact that the next interpolating distribution $\mu_{s+1}=r\big(X_s^{1:N_s}, \mu_s\big)$ is random. We remind the reader that the next interpolating distribution is chosen according to the rule:
Set $\mu_{s+1} = r\big( X_s^{1:N}, \mu_s\big) = \nu_{s, m^*}$ where
\begin{equation*}
%    \begin{split}
%        \text{Set} \; \mu_{s+1} &= r\left( X_s^{1:N}, \mu_s\right) = \nu_{s, m^*} \text{ where }\\
        %
m^* %&
= \max\big\{ m \in \left\{1,\ldots,M\right\}: \left({E}_{s,m}\geq \mathcal{E}\right)\cap \big({V}_{s,m} \geq \mathcal{C}\big) \text{ or } m=1\big\}
%    \end{split}
\end{equation*}
where $\mathcal{E}$ is a user-specified lower bound on the RESS and $\mathcal{C}$ is a tuning parameter that controls the quality of the RESS estimates. We will show that the sample size $N_s$ specified by Theorem~\ref{thm:adaptive_total_bound} \eqref{Eqn:AdaptiveSampleSize} is sufficiently large to ensure that $\B{s}$ occurs with high probability when the adaptively chosen $\mu_{s+1}$ has small $L_2$ distance from $\mu_s$ (less than $2\mathcal{E}^{-1}$). We then show that the step selection rule $r$ has a high probability of choosing such a $\mu_{s+1}$ (i.e. of taking a ``small enough" step).

\begin{lemma}\label{lemma:adaptive_concentration}
Assume~\asref{as:3}-\asref{as:5}. Suppose $\Pr(\C{s-1}) \geq 3/4$. Then for any $0<\delta_s<1$ and $t \geq \tau\big(\frac{\delta_s}{N_s}, 2\big)$ and any $0<\delta_s'<1$ and $N_s \geq \max\big( 36 \log(2M/\delta_s') \mathcal{E}^{-1}\,,\, \frac{25}{2\mathcal{C}^2} \log(4M/\delta_s')\big)$.
\begin{equation}
\label{lemma:inductive2}
    \Pr \left( \C{s} \right) \geq \left(1-\delta_s\right)\cdot \Pr\left (\C{s-1} \right)- \delta_s'.
\end{equation}
\end{lemma}
\begin{proof}
We begin by defining some mathematical objects arising in the step selection process. Let 
$$\V^*\left(\mu_s\right) = \{ \nu_{s,m} \in \V\left(\mu_s\right) : \Lp{\nu_{s,m}}{\mu_s}{2} \leq 2\mathcal{E}^{-1}\}$$ 
be the set of candidate distributions with $L_2$ distance no more than  twice the user-specified bound.  Define the event that the selected distribution lies in $\V^*\left(\mu_s\right)$:
\begin{equation}
%    \begin{split}
        \D{s} %&
        = \big\{ \mu_{s+1} \in \V^*\left(\mu_s\right) \big\}
        %\\
        %
        %&
        = \big\{ r\big(X_s^{1:N_s}, \mu_s\big) \in \V^*(\mu_s)\big\}
%    \end{split}
\end{equation}
Next we define analogous quantities and events for the coupled particles.  Let $\bar{E}_{s,m}$ be the RESS for a move to $\nu_{s,m} \in \V\left(\mu_s\right)$ computed with the coupled particles $\bar{X}_s^{1:N_s}$ where:
\begin{equation}
%    \begin{split}
        %&
        \bar{E}_{s,m} = \bar{W}_{s,m}^2 / \bar{V}_{s,m}%\\
        %&
\qquad
        \bar{W}_{s,m} = N^{-1}\sum_{i=1}^N w_{s,m}(\bar{X}_s^n)% \\
        %&
\qquad
        \bar{V}_{s,m} = N^{-1}\sum_{i=1}^N w_{s,m}^2(\bar{X}_s^n).%\\
%    \end{split}
\end{equation}
Similarly, let $\bar{\mu}_{s+1} = r\big(\bar{X}_s^{1:N_s}, \mu_s\big)$ denote the candidate distribution which would be obtained by applying the step selection procedure using the coupled particles, and define the corresponding events: 
\begin{equation}
%    \begin{split}
\BB{s} %&
= \Big{\{} \bar{W}_{s, m} > \frac{2}{3}\mu_s\left(w_{s, m}\right) \Big{\}} %\\
\qquad \quad 
\DD{s} %&
= \Big\{ \bar{\mu}_{s+1} \in \V^{*}\left(\mu_s\right) \Big\}%\\
        %
        %&
        = \Big\{ r\left(\bar{X}_s^{1:N_s}, \mu_s\right) \in \V^{*}\left(\mu_s\right)\Big\}
%    \end{split}
\end{equation}
We now prove the result. First, we show that when the probability of coupling ($\A{s}$) is high, it suffices to show that the weight concentration event for the coupled particles ($\BB{s}$) occurs with high probability:
\begin{align*}
%    \begin{split}
\Pr \left( \C{s} \right) %&
= \Pr\left( \A{s} \cap \B{s} \right) %\\
     &
     \geq \Pr\left( \A{s} \cap \B{s} \cap \D{s} \right)%\\
    %
     %&
     =\Pr\left( \A{s}  \cap \BB{s} \cap \DD{s}  \right) \\
    & 
    \geq \Pr\left( \A{s}  \right) - \left(1 - \Pr\left( \BB{s} \cap \DD{s}  \right) \right) %\\
    %
    %&
    \geq (1-\delta_s)\cdot \Pr\left(\C{s-1}\right) - \left(1 - \Pr\left( \BB{s} \cap \DD{s}  \right) \right) %\\
%    \end{split}
\end{align*}
The third line follows because the step selection $r$ chooses the same distribution when the particles are coupled, and the last line uses Lemma~\ref{lemma:coupling2}. To lower bound the probability $\Pr\left(\BB{s} \cap \DD{s} \right)$ we show that 
the size of the particle system is sufficiently large to ensure $\BB{s}$ as long as the adaptively chosen step size is not too large, i.e. as long as $\DD{s}$ also holds with high probability. To do so we define a new event:
\begin{equation}
    \BB{s,m} = \Big\{ \bar{W}_{s,m} - \mu_s\left(w_{s,m}\right) > - \mu_s\left(w_{s,m}\right)/3 \cdot \sqrt{
    %\frac{
    \Lp{\nu_{s,m}}{\mu_s}{2}
    %}{
    /
    2\mathcal{E}^{-1}}
    %}
    \Big\}.\\
\end{equation}
Recall that $m^*$ denotes the index of the selected distribution \eqref{eq:selection}, and note that when $\nu_{s, m^*} \in \V^*\left(\mu_s\right)$ then $\Lp{\mu_{s+1}}{\mu_s}{2} \leq 2 \mathcal{E}^{-1}$ and therefore the event $\BB{s, m^*} \cap \DD{s}$ contains $\BB{s}  \cap \DD{s}$. This can be used to obtain the lower bound:
\begin{align*}
%    \begin{split}
\Pr\left(\BB{s} \cap \DD{s}\right) %&
\geq  \Pr\left( \BB{s, m^*} \cap \DD{s} \right) %\\
        &\geq \Pr\left( \left\{\cap_{m\in 1:M}\; \BB{s,m} \right\}\cap \DD{s}\right) \\
        &\geq 1 - \sum_{m \in 1:M} \Pr \left( \BB{s,m}^c \right) - \Pr \left( \DD{s}^c\right)%\\
        %
        %& 
        \geq 1 - \delta_s'/2 - \Pr \left( \DD{s}^c\right).
%    \end{split}
\end{align*}
The fourth line follows from the independence of the coupled particles, the choice of $N_s$, and Bernstein's inequality (see Lemma~\ref{lemma:concentration}) which gives
\begin{equation}
    \Pr \left( \BB{s,m}^c \right) \leq \delta_s'/2M.
\end{equation}
The final step follows since $\Pr \left( \DD{s}^c\right) < \delta_s'/2$, the proof of which is deferred to the next lemma.
\end{proof}
\noindent To complete the proof of Lemma~\ref{lemma:inductive2} we 
%need to 
must show 
%that 
the step selection rule chooses a ``good'' step with high probability.  

\begin{lemma}\label{lemma:l2_selection}
Assume~\asref{as:3}-\asref{as:5} and let $\nu_{s, m^*}=r\big( \bar{X}_s^{1:N}, \mu_s\big)$.
Then for any $0<\delta_s'<1$ and $N_s \geq \frac{25}{2C^2} \log (4M/\delta_s')$:
$$\Pr \left( \DD{s}^c\right) < \delta_s'/2$$
\end{lemma}
\begin{proof}
We begin with a preliminary result. Suppose $\mu_s\left(w_{s,m}^2\right) \geq \mathcal{C}$. Then~\asref{as:3}, H\"oeffding's inequality, and this choice of $N_s$ bound the relative error of $\bar{V}_{s,m}$:
\begin{align*}
%    \begin{split}
        \Pr\big( \mu_s(w_{s,m}^2) / \bar{V}_{s,m} \geq 5/4\big)
        %
        %&= \Pr\Big( \mu_s\left(w_{s,m}^2\right)\cdot4/5 \geq \bar{V}_{\nu, s} \Big)\\
        %
        %&= \Pr\Big( \mu_s\left(w_{s,m}^2\right) -  \bar{V}_{s,m}\geq  \mu_s\left(w_{s,m}^2\right)/5\Big)\\
        %
        &= \Pr\big( \bar{V}_{s,m} - \mu_s(w_{s,m}^2)  \leq  -\mu_s(w_{s,m}^2)/5\big)\\
        &\leq \Pr\big( \bar{V}_{s,m} - \mu_s(w_{s,m}^2)  \leq  -\mathcal{C}/5\big)%\\
        %
        % &
        \leq \delta_s'/4M.%\\
%    \end{split}
\end{align*}
By~\asref{as:3} $\mu_s\left(w_{s,m}\right) \geq \mu_s\left(w_{s,m}^2\right) \geq \mathcal{C}$ and so we can also upper bound the relative error of $\bar{W}_{s,m}$:
\begin{equation}
%    \begin{split}
        \Pr\big( \bar{W}_{s,m} / \mu_s(w_{s,m})  \geq 6/5\big) 
        %
        %&=\Pr\Big( \bar{W}_{\nu, s}  \geq \mu_s\left(w_{\nu, s}\right) \cdot 6/5\Big)\\
        %
        %&=\Pr\Big( \bar{W}_{s,m} - \mu_s\left(w_{s,m}\right)  \geq \mu_s\left(w_{s,m}\right)/5\Big)\\
        %
        %&
        \leq \Pr\big( \bar{W}_{s,m} - \mu_s(w_{s,m})  \geq \mathcal{C}/5\big)%\\
        %
        %&
        \leq \delta_s'/4M.\\
%    \end{split}
\end{equation}
Together these provide an upper bound on the relative error of the RESS as an estimate of the $L_2$ distance when $\mu_s\left(w_{s,m}^2\right) \geq \mathcal{C}$:
\begin{equation}
%    \begin{split}
        \Pr \Big( \frac{\bar{E}_{s,m}}{\Lp{\nu_{s,m}}{\mu_s}{2}} < \frac{(6/5)^2}{4/5} \Big) 
        %
        %&= \Pr \left( \frac{\bar{W}_{s,m}^2}{\bar{V}_{s,m}} \cdot \frac{\mu_s\left(w_{s,m}^2\right)}{\mu_s\left(w_{s,m}\right)^2} < (6/5)^2 \cdot 5/4 \right) \\
        %
        %&= \Pr \left( \frac{\bar{W}_{s,m}^2}{\mu_s\left(w_{s,m}\right)^2} \cdot \frac{\bar{V}_{s,m}}{\mu_s\left(w_{s,m}^2\right)} < (6/5)^2 \cdot 5/4 \right) \\
        %
        %&
        \geq \Pr \Big( \Big\{\frac{\mu_s(w_{s,m}^2)}{\bar{V}_{s,m}} < 5/4\Big\} \cap \Big\{\frac{\bar{W}_{s,m}}{\mu_s(w_{s,m})} < 6/5 \Big\} \Big)%\\
        %
        %&
        \geq 1-\delta_s'/2M\\
%    \end{split}
\end{equation}
using an intersection bound. When %we have
both $\mu_s\left(w_{s,m}^2\right) \geq \mathcal{C}$ and $\Lp{\nu_{s,m}}{\mu_s}{2} > 2\mathcal{E}^{-1}$ this gives:
$$\Pr \left( \bar{E}_{s,m} \geq \mathcal{E}\right) \leq \delta_s'/2M.$$
In the case that $\mu_s\left(w_{s,m}^2\right) < \mathcal{C}$,~\asref{as:3}, H\"oeffding's inequality, and the choice of $N_s$ ensure:
\begin{equation*}
%    \begin{split}
\Pr \left( \bar{V}_{s,m} - \mu_s\left(w_{s,m}^2\right) \geq \mathcal{C}/5 \right) %&
\leq \Pr \left( \bar{V}_{s,m} - \mathcal{C} \geq C/5 \right)
% \\
        %
        %&
        \leq \Pr \left( \bar{V}_{s,m} \geq 6/5\cdot \mathcal{C} \right)
        %\\
        %
        %&
        \leq  \delta_s'/2M%\\
%    \end{split}
\end{equation*}
We now show the result. Let:
\begin{equation}
    \begin{split}
        Q\left(\mu_s\right) = \left\{ \nu_{s,m} \in \V\left(\mu_s\right) : \big(\Lp{\nu_{s,m}}{\mu_s}{2} > 2\mathcal{E}^{-1} \big) \cap \big( \mu_s\left(w_{s,m}^2\right) \geq \mathcal{C} \big)\right\} \\
        R\left(\mu_s\right) = \left\{  \nu_{s,m} \in \V\left(\mu_s\right) : \big(\Lp{\nu_{s,m}}{\mu_s}{2} > 2\mathcal{E}^{-1} \big) \cap \big( \mu_s\left(w_{s,m}^2\right) < \mathcal{C} \big)\right\}
    \end{split}
\end{equation}
We upper bound the probability of selecting the next candidate distribution from $\V\left(\mu_s\right) \setminus \V^*\left(\mu_s\right)$.
\begin{align*}
%    \begin{split}
\Pr \left( \DD{s}^c\right) 
&= \Pr\big( r(\bar{X}_s^{1:N_s}, \mu_s) \notin \V^*(\mu_s)\big)
    %\\
        %
        %&
\leq \Pr \big( \cup_{\nu_{s,m} \notin \V^*(\mu_s)} (\bar{E}_{s,m} \geq \mathcal{E}) \cap (\bar{V}_{s,m} \geq 6/5 \cdot \mathcal{C}) \big)\\
        %
        %&\scscancel{\leq \sum_{\nu \notin \V_{\mu_s}^*} \Pr \left( \big(\bar{E}_{\nu,s} \geq \mathcal{E}\big) \cap \big(\bar{V}_{\nu, s} \geq 6/5 \cdot \mathcal{C} \big) \right)} \\
        %
        &\leq \sum_{\nu_{s,m} \in Q(\mu_s)} \Pr \big((\bar{E}_{s,m} \geq \mathcal{E}) \cap (\bar{V}_{s,m} \geq 6/5 \cdot \mathcal{C})\big) +
       % \\
        %
        %&\quad\; 
        \sum_{\nu_{s,m} \in R(\mu_s)} \Pr \big((\bar{E}_{s,m} \geq \mathcal{E}) \cap (\bar{V}_{s,m} \geq 6/5 \cdot \mathcal{C}) \big)\\
        &\leq \sum_{\nu_{s,m} \in Q(\mu_s)} \Pr \big( \bar{E}_{s,m} \geq \mathcal{E}\big) + \sum_{\nu_{s,m} \in R(\mu_s)} \Pr \big( \bar{V}_{s,m} \geq 6/5 \cdot \mathcal{C} \big)\\
        &\leq \sum_{\nu_{s,m} \in Q(\mu_s)} \delta_s' / 2M + \sum_{\nu_{s,m} \in R(\mu_s)}  \delta_s' / 2M
        %\\
        %
        %&
        \leq \delta_s' / 2%\\
%    \end{split}
\end{align*}
where the second to last line uses the preliminary results above.

\end{proof}
Lemma~\ref{lemma:l2_selection}, combined with the coupling argument from Lemma~\ref{lemma:adaptive_concentration}, guarantees %that 
the selected $\mu_{s+1}$ will, with high probability, satisfy the user-specified upper bound on step-size, but it does not guarantee that the selected stepsize will be close to this bound; $\Lp{\mu_{s+1}}{\mu_s}{2}$ may be much less than $\mathcal{E}^{-1}$, in which case the algorithm may take shorter steps than desired, resulting in a longer than necessary path. 

The proof of Theorem~\ref{thm:adaptive_total_bound} is completed by applying Corollary~\ref{lemma:inductive2} inductively to establish that $\A{S}$ holds with high probability. Due to the adaptively selected stepsize, the number of steps $S$ is random and so $\A{S}$ must hold for any $S$ at which the algorithm terminates.

\begin{proof}[Proof of Theorem~\ref{thm:adaptive_total_bound}]
The algorithm terminates when $\pi$ is chosen as the candidate distribution. Let $S$ be the step at which this occurs; $S$ is finite by~\asref{as:5}. As shown in Theorem 1 of ~\cite{Marion2018}, the error of SMC estimator can be lower bounded by:
\begin{equation}
    \begin{split}
        \Pr\left( |\hat{f} - \pi(f)|\leq \epsilon\right)     &\geq \left(1-\delta_{S}\right)\cdot \Pr\left (\C{S-1} \right)- \delta_{S}'.
    \end{split}
\end{equation}
For any $S$, $\Pr\left (\C{S-1} \right)$ can be lower bounded using Lemma~\ref{lemma:inductive2} inductively.  We sill show that for $s\geq1$:
\begin{equation}
    \Pr\left(\C{s}\right) \geq \prod_{r=1}^s(1-\delta_r) - \sum_{q=0}^{s-1} \delta'_q \Big( \prod_{r=q+1}^s(1-\delta_r)\Big) - \delta'_s.
\end{equation}
The proof follows by induction.  The base case is established by noting that $\Pr\left(\C{0}\right) \geq 1 - \delta_0'$ since $\A{0}$ holds by definition and $\B{0}$ follows from Bernstein's inequality. Therefore by Lemma~\ref{lemma:inductive2}:
\begin{equation*}
\Pr\left(\C{1}\right) \geq \left(1-\delta_1\right)\cdot\left( 1 - \delta_0'\right) -\delta_1'%\\
        %
        %&= 
        \left(1-\delta_1\right) - \delta_0'\left(1-\delta_1\right) - \delta'_1.
\end{equation*}
To show the inductive step assume that the statement holds for step $s-1$. Then
\begin{align*}
%    \begin{split}
\Pr \left( \C{s} \right) %&
\geq \left(1-\delta_s\right)\cdot \Pr\left (\C{s-1} \right)- \delta_s'%\\
        &\geq \left(1-\delta_s\right)\cdot \Big( \prod_{r=1}^{s-1}(1-\delta_r) - \sum_{q=0}^{s-2} \delta'_q \Big( \prod_{r=q+1}^{s-1}(1-\delta_r)\Big) - \delta'_{s-1}\Big) - \delta_s'\\
        &=\prod_{r=1}^s(1-\delta_r)  - \sum_{q=0}^{s-1} \delta'_q \Big( \prod_{r=q+1}^s(1-\delta_r)\Big) - \delta'_s.%\\
%    \end{split}
\end{align*}
The second line holds by Lemma~\ref{lemma:inductive2}.  Having shown the induction holds, we use it to obtain a lower bound. For any $S\geq 1:$
\begin{align*}
 %   \begin{split}
\Pr\big( |\hat{f} - \pi(f)|\leq \epsilon\big)     &\geq \left(1-\delta_{S}\right)\cdot \Pr\left (\C{S-1} \right)- \delta_{S}'\\
        &
\geq\prod_{r=1}^{S}\left(1-\delta_r\right)  - \sum_{q=0}^{S-1} \delta'_q \Big( \prod_{r=q+1}^{S}\left(1-\delta_r\right)\Big) - \delta'_{S}%\\
        %
        % &\scsxcancel{\geq\prod_{r=1}^{S}\left(1-\delta_r\right)  - \sum_{q=0}^{S-1} \delta'_q \left( \prod_{r=1}^{S}\left(1-\delta_r\right)\right) - \delta'_{S}}\\
        %
        % &\scsxcancel{\geq\prod_{r=1}^{S}\left(1-\delta_r\right)  - \sum_{q=0}^{S} \delta'_q \left( \prod_{r=1}^{S}\left(1-\delta_r\right)\right)}
        %
%&
\geq\Big(\prod_{r=1}^{S}\left(1-\delta_r\right)\Big) \cdot \Big(1 - \sum_{q=0}^S \delta'_{q}\Big).%\\
%    \end{split}
\end{align*}
To complete the proof we choose a sequence for $\delta_s$ and $\delta_s'$ so that this product is lower bounded by $3/4$ for any $S\geq1$.   Choosing $\delta_s=\left(4 s\right)^{-2}$ gives:
\begin{equation}
    \prod_{s=1}^{S}\left(1-\delta_s\right) > %\scscancel{\prod_{s=1}^{\infty}\left(1-\delta_s\right)}
        %
       % \scscancel{=}
       \prod_{s=1}^{\infty}\left(1-\left(4 s\right)^{-2}\right)
        = \frac{\sin \left(\pi/4\right)}{\pi/4}.
\end{equation}
The lower bound is from the infinite product representation of the sinc function, attributed to Euler. Choosing $\delta_s'=\left(1+s\right)^{-2}/10$ gives:
\begin{equation}
        \sum_{q=0}^S \delta'_s < % \scscancel{\sum_{q=0}^\infty \delta'_s}
        %
        %\scscancel{=} 
        \frac{1}{10} \sum_{s=0}^\infty (1+s)^{-2}
        = \frac{\pi^2}{60}.
\end{equation}
This equality is the solution to the Basel problem.  Combining these results gives:
\begin{equation}
\Pr\big( |\hat{f} - \pi(f)|\leq \epsilon\big) \geq\Big(\prod_{r=1}^{S}\left(1-\delta_r\Big)\right) \cdot \Big(1 - \sum_{q=0}^S \delta'_{q}\Big)
\geq \frac{\sin (\pi/4)}{\pi/4} \cdot \big(1-\frac{\pi^2}{60}\big)
> 3/4.
\end{equation}
\end{proof}
The choice of sequences $(\delta_s)_{s=1}^S$ and $(\delta'_s)_{s=1}^S$ are somewhat arbitrary; 
%choices of the proof. 
other %selections 
choices would affect the upper bounds on $N_s$ and $t_s$ provided in Theorem~\ref{thm:adaptive_total_bound}.  This choice determines the tradeoff between $N_s$ and $t_s$ and can also affect how the cost of the algorithm grows with $s$.  For example, choosing $\delta_s$ %so that it starts
to start relatively large and go to zero rapidly results in an algorithm with a number of particles which starts small ($N_0$) but increases rapidly as $s$ increases.  Alternatively, choosing $\delta_s$ to be small initially and to decay more slowly results in a larger number of initial particles that grows more slowly with $s$.  In either case, the complexity of the algorithm grows logarithmically in $1/\delta_s$ and $1/\delta_s'$, and so for any convergent sequence the complexity of our bound will be no better than $N_s = \OS{\log s}$ and $t_s = \OS{\log s}$.  

Note %that 
the complexity of the adaptive algorithm is comparable to that
%the cost 
of the non-adaptive algorithm from~\cite{Marion2018} using the same %sequence of distributions,
distribution sequence, 
as measured by the total number of steps $S$ (%for which 
the non-adaptive algorithm requires $N = \OS{\log S}$ and $t = \OS{\log S}$).

\subsection{Theorem~\ref{thm:another_adaptive_total_bound}}
\label{Sec:AltThm2}
This section contains an alternate version of Theorem~\ref{thm:adaptive_total_bound} that uses a modified version of the step selection rule in equation~\eqref{eq:selection} that removes the `$\mathcal{C}$' condition: Set $\mu_{s+1} = r'( X_s^{1:N}, \mu_s) = \nu_{s, m^*}$ where
\begin{equation}
\label{eq:selection2}
%    \begin{split}
%        \text{Set} \; \mu_{s+1} &= r'\left( X_s^{1:N}, \mu_s\right) = \nu_{s, m^*} \text{ where }\\
        %
        m^* %&
        = \max\left\{ m \in \left\{1,\ldots,M\right\}: {E}_{s,m}\geq \mathcal{E} \text{ or } m=1\right\}.
%    \end{split}
\end{equation}
This rule more closely matches how adaptive, RESS guided SMC is performed in practice.  In the absence of the $\mathcal{C}$ condition, the bounds in the following Theorem depend instead on the maximal $L_4$ distance between candidate pairs
$$ \max_{\mu \in \V,\, \nu \in \V(\mu)} \Lp{\nu}{\mu}{4}.$$
The $L_4$ distance is used to apply Lemma~\ref{Appendix E: ESS bound}, which quantifies the relative error of the ESS as an estimate of the $L_2$ distance.
\begin{theorem}[Error Bound for Adaptive Step-Selection SMC Using Selection Rule $r'$]
\label{thm:another_adaptive_total_bound} Choose  $\mathcal{E}\in\left(0, 0.5\right)$ and assume~\asref{as:3}-\asref{as:5}.  For $s = 0,1,\ldots,S$ set
\begin{equation}
N_s = \max 
\begin{cases}
36 \cdot \gamma(s) \cdot \mathcal{E}^{-1}\\
50 \cdot \Big(\gamma(s) + \log(4)\Big) \cdot \Big( \max_{\mu \in \V,\, \nu \in \V(\mu)} \Lp{\nu}{\mu}{4}\Big) \\
1/2 \cdot \gamma(s) \cdot \epsilon^{-2}
\end{cases}
\end{equation}
for $\gamma(s) = \log(20M(1+s^2))$. Let $\tau\left(\;\cdot\;, 2\right) = \sup_{\mu \in \V} \tau_\mu\left(\;\cdot\;, 2\right)$ and for $s\geq1$ set
\begin{equation}
t_s \geq \tau\big( (16 s^ 2N_s)^{-1},\; 2\big). 
\end{equation}
Fix $\epsilon > 0$ and draw $X_0^{1:N_0}\iidsim\mu_0$. Then for any $f\in\mathcal{F}$ with $|f|\leq 1$ the adaptive SMC algorithm ensures $\big|\hat f - \pi \left(f\right)\big| \leq \epsilon$ with probability at least $3/4$.
\end{theorem}
The proof of this theorem is obtained by replacing Lemma~\ref{lemma:l2_selection} with the following:
\begin{lemma}
\label{lemma:l2_selection_redo}
Assume~\asref{as:3}-\asref{as:5} and let $\nu_{s, m^*}=r'\big( \bar{X}_s^{1:N}, \mu_s\big)$.
Then for any $0<\delta_s'<1$ and $N_s \geq 50 \log \big(\frac{8M}{\delta_s'} \big) \cdot \big( \max_{\nu_{m,s} \in \V(\mu_s)} \Lp{\nu_{m,s}}{\mu_s}{4}\big)$:
$$\Pr \left( \DD{s}^c\right) < \delta_s'/2$$
\end{lemma}
\begin{proof}
We remind the reader of the following definition.
$$\V^*\left(\mu_s\right) = \{ \nu_{s,m} \in \V\left(\mu_s\right) : \Lp{\nu_{s,m}}{\mu_s}{2} \leq 2\mathcal{E}^{-1}\}$$ 
To prove the result we will show that $\nu_{s,m} \notin \mathcal{V}^*\left(\mu_s\right)$ implies  $\Pr\left( \bar{E}_{s,m} \geq \mathcal{E}\right) < \delta_s'/M$. Applying Lemma~\ref{Appendix E: ESS bound} with $\epsilon=1/5$ and $\delta = \delta_s'/2M$ gives
\begin{equation}
\Pr \left( \bar{E}_{s,m} \cdot \Lp{\nu_{s,m}}{\mu_s}{2} \in \left[0.5, 2\right]\right) \geq  1-\delta_s'/2M. \end{equation}
The result follows:
\begin{align*}
%    \begin{split}
\Pr \left( \DD{s}^c\right) 
        %
        %&
        = \Pr\big( r(\bar{X}_s^{1:N_s}, \mu_s) \notin \V^*(\mu_s)\big)%\\
        %
        %&
        \leq \Pr( \cup_{\nu_{s,m} \notin \V^*(\mu_s)} \{\bar{E}_{s,m} \geq \mathcal{E} \})%\\
        %
        %&
        \leq \sum_{{\nu_{s,m} \notin \V^*\left(\mu_s\right)}}  \Pr ( \bar{E}_{s,m} \geq \mathcal{E} )%\\
        %
        %&
        \leq \delta_s' / 2.%\\
%    \end{split}
\end{align*}
\end{proof}
\noindent The advantage of Theorem~\ref{thm:another_adaptive_total_bound} is that it applies directly to the unmodified selection rule $r'$ that is commonly used in practice.  An implication of Lemma~\ref{lemma:l2_selection_redo} is that, with high probability, the estimates of the $L_2$ distances used in step selection have a relative error of Bounded above 2 and below by $\frac{1}{2}$.  Therefore, with high probability each step in the distribution sequence chosen by the algorithm has $L_2$ distance lying in  $[0.5\mathcal{E}^{-1}, 2\mathcal{E}^{-1}]$.  This is in contrast to Theorem~\ref{thm:adaptive_total_bound}, which makes does not lower-bound the stepsize of the chosen path.
  
The limitation of Theorem~\ref{thm:another_adaptive_total_bound} is that the reliance on the worst-case $L_4$ distance makes applying the bound impractical.  In cases where the worst-case $L_4$ distance can be bounded, the $L_2$ distance to all candidate distributions is also bounded and Theorem~\ref{thm:total_bound} can be applied using a non-adaptive path that simply takes the largest jump at each step of the algorithm. Introduction of the $\mathcal{C}$ condition in rule $r$ alleviates/avoids this reliance on an $L_4$ bound, at the expense of modifying the standard selection rule.

\section{Examples}

\subsection{Gaussian example}\label{sec:appendix_gaussian}
First, we derive the $L_2$ distance from $\eta \sim N(\theta_0, \phi_0^{-1})$ to $\mu \sim N(\theta_1, \phi_1^{-1})$ %on $\mathbb{R}$, 
assuming $2\phi_1\geq \phi_0$. We have
\begin{equation}\label{eq:l2_part1}
    \begin{split}
        \| \mu/\eta \|_{L_2(\eta)} = \frac{\phi_1}{\phi_0^{1/2}}\int \frac{1}{\sqrt{2\pi}}\exp\big(-0.5\big[2\phi_1(x-\theta_1)^2 - \phi_0(x-\theta_0)^2\big]\big)
    \end{split}
\end{equation}
Let $\phi^* = 2\phi_1-\phi_0$ and $\theta^* = 2\phi_1\theta_1-\phi_0\theta_0$ and complete the square inside the exponential function. 
\begin{equation}
\label{eq:l2_part2}
%    \begin{split}
        2\phi_1(x-\theta_1)^2-\phi_0(x-\theta_0)^2
        %
        %&=x^2\phi^* -2x\phi^* \frac{\theta^*}{\phi^*}+\frac{{\theta^*}^2}{\phi^*}-\frac{{\theta^*}^2}{\phi^*}+2\phi_1\theta_1^2 - \phi_0\theta_0^2\\
        %
        %&
        = \phi^*\big(x-\theta^*/\phi^*\big)^2+2\phi_1\theta_1^2 - \phi_0\theta_0^2-\frac{{\theta^*}^2}{\phi^*}%\\
        %
        %&
        = \phi^*\big(x-\theta^*/\phi^*\big)^2 - \frac{2\phi_1\phi_0}{\phi^*}(\theta_1 - \theta_0)^2
%    \end{split}
\end{equation}
Inserting~\eqref{eq:l2_part2} into~\eqref{eq:l2_part1} and substituting $\psi = \phi_1/\phi_0$ gives
\begin{equation}
%\label{eq:l2_part3}  I don't think this is referenced
    \begin{split}
        \| \mu/\eta \|_{L_2(\eta)} = \frac{\psi}{\sqrt{2\psi-1}}\exp\Big(\frac{\phi_1}{2\psi -1}(\theta_1 - \theta_0)^2\Big)
    \end{split}
\end{equation}
The $L_2$ distance for spherical, $d$-dimensional Gaussians follows immediately:
\begin{equation}\label{eq:L2_gaussian}
  \| \mu/\eta \|_{L_2(\eta)} = \Big(\frac{\psi^2}{2\psi-1}\Big)^{d/2}\exp\Big(\frac{d\phi_1}{2\psi -1}(\theta_1 - \theta_0)^2\Big) 
\end{equation}
\subsection{Geometric path}
\label{Sec:GaussianGeometricPath}
The geometric path from section~\ref{sec:gaussian} consists of a sequence of Gaussian distributions with $\mu_{\beta_s}(x) = N\big(1_d\cdot \theta_s\,,\, I_d /\phi_s\big)$ where $\theta_s = \theta\cdot \beta_s/\phi_s$ and $\phi_s=\beta_s(\phi-1) +1$. We remind the reader that $\phi > 1$ and $\theta\geq 2$ and proceed to bound the $L_2$ distance by separately bounding the factors in~\eqref{eq:L2_gaussian}. Define $\psi_s = \phi_s/\phi_{s-1}$. For $s=1$:d
\begin{equation*}
%    \begin{split}
1 < \psi_1 %&
= 1 + \frac{\phi-1}{\phi\cdot\theta\sqrt{d}} %\\
        %
        %& 
\leq  1 + \frac{2}{\theta\sqrt{d}}
%    \end{split}
\end{equation*}
and when $s>1$:
\begin{equation*}
%\begin{split}
1\leq \psi_s 
        %&
        = \frac{\beta_s(\phi-1)+1}{\beta_{s-1}(\phi-1)+1}%\\
        %
        %&
        = 1 + \frac{(\beta_s - \beta_{s-1})(\phi-1)}{\beta_{s-1}(\phi-1)+1}%\\
        %
        %&
        \leq1 + \frac{(\beta_s - \beta_{s-1})}{\beta_{s-1}}%\\
        %&
        =1 + \frac{2}{\theta\sqrt{d}}
%    \end{split}
\end{equation*}
Plugging this into the factor $\Big(\frac{\psi^2}{2\psi-1}\Big)^{d/2}$in the $L_2$ distance \eqref{eq:L2_gaussian} gives
\begin{equation}
\label{eq:gaussian_temper1}
%    \begin{split}
\Big(\frac{\psi_s^2}{2\psi_s-1}\Big)^{d/2} 
        %& 
        \leq 
        %\Bigg(\frac{\big(1 + \frac{2}{\theta\sqrt{d}}\big)^2}{2\big(1 + \frac{2}{\theta\sqrt{d}}\big)-1}\Bigg)^{d/2}\\
        %& =
        \Big(\frac{\big(1 + \frac{2}{\theta\sqrt{d}}\big)^2}{\big(1 + \frac{4}{\theta\sqrt{d}}\big)}\Big)^{d/2} %\\
        %
        %&
        \leq \Big(1+\frac{1}{d}\Big)^{d/2}%\\
        %
        %&
        \leq 2%\\
%    \end{split}
\end{equation}
where the second line uses $\theta\geq 2$. To bound the second factor in \eqref{eq:L2_gaussian} we bound the difference in means. For $s=1$, $\theta_1 - \theta_0 = \frac{1}{\phi\sqrt{d} \cdot \phi_1} \leq \frac{1}{\sqrt{d \phi_1} }$ and consequently $\exp\Big(\frac{d\phi_1}{2\psi_1 -1}(\theta_1 - \theta_0)^2\Big) \leq e$. For $s>1$:
%.
\begin{equation*}
\label{eq:gaussian_temper2}
%    \begin{split}
\theta_s - \theta_{s-1}
        %
        %&
        =\theta\Big(\frac{\beta_s}{\phi_s}-\frac{\beta_{s-1}}{\phi_{s-1}}\Big)%\\
        %
        %&
        =\frac{\theta \cdot \beta_{s-1}}{\phi_{s-1}\phi_s} \Big(\big(1+\frac{2}{\theta\sqrt{d}}\big)\phi_{s-1} - \phi_s\Big)%\\
        %
        %&
        = \frac{  2\beta_{s-1}}{\phi_{s-1}\phi_s\sqrt{d}} %\\
%    \end{split}
\end{equation*}
Inserting this result into the second term in~\eqref{eq:L2_gaussian} gives
\begin{equation*}
%    \begin{split}
 \exp\Big(\frac{d\phi_1}{2\psi -1}(\theta_1 - \theta_0)^2\Big) 
        %
        %&
        = \exp\Big(\frac{4\beta_{s-1}^2}{2\phi_s^2\phi_{s-1} - \phi_s \phi_{s-1}^2}\Big)\\
        %
        %&
        \leq \exp(4)
%    \end{split}
\end{equation*}
The first line follows using $1\leq \phi_{s-1}\leq \phi_s$ and $\beta_s\leq 1$. Inserting~\eqref{eq:gaussian_temper1} and~\eqref{eq:gaussian_temper2} into~\eqref{eq:L2_gaussian} shows that for the geometric path $1/\mathcal{E} = \bigO(1)$ proving~\eqref{eq:tempered_gaussian}. 
\subsection{Precision path}\label{sec:GaussianPrecisionPath}The precision path is specified by a sequence of normal distributions $\mu_s = N_d(\theta_s, I_d/\phi_s)$. The location parameter is $\theta_s = 0$ for $s\leq s_1= \ceil{3\sqrt{d}\log(d\theta^2)}$ and $\theta_s = 1_d\theta$ otherwise. The precisions are given by
\begin{equation}
    \phi_s = 
    \begin{cases}
        \big(1-\frac{1}{\sqrt{9d}}\big)^s  \vee \frac{1}{d\theta^2}, & \text{if}\ 0\leq s\leq s_1 \\
        \frac{1}{d\theta^2}\big(1+\frac{1}{\sqrt{d}}\big)^{s-s_1-1}\wedge \phi, & \text{otherwise}
    \end{cases}
\end{equation}
Let $\psi_s = \phi_s/\phi_{s-1}$. When $s\leq s_1$, $1\geq \psi_s \geq \big(1-\frac{1}{\sqrt{9d}}\big)$ and therefore
\begin{equation}
\label{eq:precisian_gaussian1}
%    \begin{split}
 \|\mu_s\|_{L_2)(\mu_{s-1})} 
        %
        %&
        = \Big(\frac{\psi_s^2}{2\psi_s-1}\Big)^{d/2}%\\
        %
        %&
        \leq \Big(\frac{\big(1 - \frac{1}{\sqrt{9d}}\big)^2}{\big(1 - \frac{2}{\sqrt{9d}}\big)}\Big)^{d/2}% \\
        %
        %&
        \leq \big(1+\frac{1}{d}\big)^{d/2}%\\
        %
        %&
        \leq 2%\\
%    \end{split}
\end{equation}
The same approach shows that $\|\mu_s/\mu_{s-1}\|_{L_2(\mu_{s-1})}\leq 2$ for $s\geq s_1+2$. When $s=s_1 + 1$, $\phi_s =1$ and therefore $\|\mu_s/\mu_{s-1}\|_{L_2(\mu_{s-1})} = 1$ and therefore $1/\mathcal{E}\leq 2$, proving~\eqref{eq:precision_gaussian}.

\subsection{Bayesian linear regression example}\label{sec:regression_theory}
The specified Bayesian linear model leads to a Normal Inverse-Gamma posterior distribution with:
$$\mu_s(\beta,\sigma^2\mid X_{1:k_s}, Y_{1:k_s}) = \mathcal{N}(\beta \mid \theta_s, \sigma^2_s \Sigma_s)\cdot \text{Inv-Gamma}(\sigma^2 \mid a_s, b_s)$$
%
%where:
\begin{align*}
%    \begin{split}
\text{ where } \qquad \Sigma_s &= \big(\Sigma_0 + X_{1:k_s}^T X_{1:k_s}\big)^{-1}&
%\qquad\qquad 
a_s &= 4 + k_s/2&\\
\theta_s &= \Sigma_s X_{1:k_s}^T Y_{1:k_s}&
%\qquad \qquad 
 b_s &= 4 + \frac{1}{2}\big(Y_{1:k_s}^T Y_{1:k_s} - \theta_s^T \Sigma_S^{-1} \theta_s\big)&
 %   \end{split}
\end{align*}
and $\Sigma_0 = (X_{1:K}^TX_{1:K})^{-1}/K$. The $L_2$ distance is:
\begin{equation}
    \begin{split}
        \|\mu_s/\mu_{s-1}\|_{L_2(\mu_{s-1})} &= \int \frac{\mathcal{N}^2(\beta \mid \theta_{s-1}, \sigma^2 \Sigma_{s-1}) \cdot \text{Inv-gamma}^2(\sigma^2 \mid a_{s-1}, b_{s-1})}{\mathcal{N}(\beta \mid \theta_s, \sigma^2 \Sigma_s) \cdot \text{Inv-gamma}(\sigma^2 \mid a_s, b_s)}d\beta d\sigma^2\\
    \end{split}
\end{equation}
The conditional normal distribution on $\beta$ can be integrated out by completing the square:
\begin{align*}
 %   \begin{split}
\int \frac{\mathcal{N}^2(\beta \mid \theta_{s-1}, \sigma^2 \Sigma_{s-1}) }{\mathcal{N}(\beta \mid \theta_s, \sigma^2 \Sigma_s)}d\beta &=\int \frac{|2\pi\sigma^2\Sigma_{s-1}|^{-1} }{|2\pi\sigma^2\Sigma_s|^{-1/2} }%\cdot\\
        %
        %&\quad \quad 
        \frac{\exp\big(-\frac{1}{\sigma^2}(\beta-\theta_{s-1})^T \Sigma_{s-1}^{-1}(\beta-\theta_{s-1})\big)}{\exp\big(-\frac{1}{2\sigma^2}(\beta-\theta_s)^T \Sigma_s^{-1}(\beta-\theta_s)\big)}d\beta\\
        &= \frac{|\Sigma_s|^{1/2} | \Sigma_*|^{1/2}}{|\Sigma_{s-1}|}\exp\Big(-\frac{b_*}{\sigma^2}\Big)%\\
%    \end{split}
\end{align*}
where $\Sigma_* =\big(2\Sigma_{s-1}^{-1} - \Sigma_s^{-1}\big)^{-1}$, $\mu_* = \Sigma_* \big(2\Sigma_{s-1}^{-1}\theta_{s-1} - \Sigma_s^{-1}\theta_s\big)$, and $b_* = \frac{1}{2} \big[ 2 \theta_{s-1}^T \Sigma_{s-1}^{-1}\theta_{s-1} - \theta_s^T \Sigma_s^{-1}\theta_s - \theta_{*}^T \Sigma_{*}^{-1}\theta_{*}\big]$. The $L_2$ distance can be found by integrating the resulting unnormalized gamma pdf:
\begin{align*}
%    \begin{split}
    \|\mu_s/\mu_{s-1}\|_{L_2(\mu_{s-1})} &= \int \frac{\text{Inv-gamma}^2(\sigma^2 \mid a_{s-1}, b_{s-1})}{\text{Inv-gamma}(\sigma^2 \mid a_s, b_s)}\cdot%\\
    %
    %&
    %\quad\quad
    \frac{|\Sigma_s|^{1/2} | \Sigma_*|^{1/2}}{|\Sigma_{s-1}|}\exp\bigg(-\frac{b_*}{\sigma^2}\bigg) d\sigma^2\\
    &= \frac{|\Sigma_s|^{1/2} | \Sigma_*|^{1/2}}{|\Sigma_{s-1}|} \cdot \frac{b_{s-1}^{2a_{s-1}} }{b_s^{a_s}(b_*+2b_{s-1}-b_s)^{2a_{s-1} - a_s}} \cdot%\\
    %
%&
%\quad\quad
\frac{\Gamma(a_s)\Gamma(2a_{s-1}-a_s)}{\Gamma(a_{s-1})^2}
%    \end{split}
\end{align*}

\section{Concentration Results}\label{sec: Concentration Results}
\begin{lemma}\label{Appendix E: L2 Bound}
Suppose $0 < w_s(X^i_{s-1}) < 1$, where $X^{1:N}_{s-1} \sim \mu_{s-1}$ are independent and identically distributed. Then for $0 < \delta < 1$,
\begin{equation}
    \Pr\Big(\frac{1}{N}\sum^N_{i=1} w_s(X^i_{s-1}) \geq \mu_{s-1}(w_s) \cdot \frac{2}{3} \Big) > 1 - \delta,
\end{equation}
for $N > 18 \log(\frac{1}{\delta}) \cdot \mathcal{E}^{-1}$
\end{lemma}
\begin{proof}
Let $X^{1},\ldots,X^N$ be independent and identically distributed random variables such that $X^i \leq b$. Then the one-sided Bernstein inequality is given by
\begin{equation}
\Pr\Big(\frac{1}{N}\sum^N_{i=1} X_i - \E[X_i] \geq t \Big) \leq \exp\Big\{-\frac{N t^2}{2\big(\E[X^2_i] + \frac{b t}{3} \big)} \Big\}
\end{equation}
Note that $ \|\mu_s/\mu_{s-1}\|_{L_2(\mu_{s-1})} = \frac{\mu_{s-1}(w^2_s)}{[\mu_{s-1}(w_s)]^2}$. Applying Bernstein's inequality to $-w_s(X^i_{s-1}) < 0$, we get
\begin{equation*}
%    \begin{split}
\Pr\Big(\frac{1}{N}\sum^N_{i=1} w_s(X^i_{s-1}) \leq \mu_{s-1}(w_s) \cdot \frac{2}{3} \Big) %&
\leq \exp\Big\{-\frac{N}{18  \|\mu_s/\mu_{s-1}\|_{L_2(\mu_{s-1})} } \Big\} %\\
        %
        %&
\leq \exp\Big\{-\frac{N \mathcal{E}}{18 } \Big\} 
%    \end{split}
\end{equation*}
Setting $N > 18 \log(\frac{1}{\delta}) / \mathcal{E}$ gives the result.
\end{proof}

We can actually make a stronger statement since $w_{x}(X_s) < 1$.

\begin{lemma}\label{Appendix E: weight bound}
Suppose $0 < w_s(X^i_{s-1}) < 1$ and $X^{1:N}_{s-1} \sim \mu_{s-1}$ are independent and identically distributed. Let $0 < \epsilon < 1$ and $0 < \delta < 1$. Then
\begin{equation}
    \Pr\Big(\Big|\frac{1}{N}\sum^N_{i=1} w^p_s(X^i_{s-1}) - \mu_{s-1}(w^p_s) \Big| \geq  \epsilon \cdot \mu_{s-1}(w^p_s) \Big) \leq \delta
\end{equation}
for $N > \frac{2}{\epsilon^2} \cdot \log\left(\frac{2}{\delta}\right) \cdot  \norm{\frac{\mu_s}{\mu_{s-1}}}_{L_{2p}(\mu_{s-1})}$.
\end{lemma}

\begin{proof} 
We first apply the following result from \cite{bercu2015}. Let $X_1,\ldots,X_N$ be a sequence of independent centered random variables. Let $\Var[X_k] \leq v_k$ and $X_k \leq b_k$. Then
\begin{equation}
    \Pr\Big(\sum^N_{i=1} X_i  \geq t \Big) \leq \exp\Big\{-\frac{3t^2}{6V_N + B_N} \Big\},
\end{equation}
where $V_N = \sum^N_{k=1} v_k$ and $B_N = \sum^N_{k=1} \big(b_k -\frac{v_k}{b_k}\big)^2_+$, with $(\cdot)_+ = \max\{0,\cdot \}$. Set $b_k = 1$ and $v_k = \Var[w^p_s(X_k)]$. Since $\Var[w^p_s(X_i)] \leq 1$, $v_k \leq b_k = 1$ and so $B_N = 0$. It follows that
\begin{equation}
    \Pr\Big(\frac{1}{N}\sum^N_{i=1} w^p_s(X^i_{s-1})  \geq (1 + \epsilon) \cdot \mu_{s-1}(w^p_s)  \Big) \leq \exp\Big\{-\frac{N \epsilon^2}{2 \norm{\frac{\mu_s}{\mu_{s-1}}}_{L_{2p}(\mu_{s-1})}} \Big\},
\end{equation}
where we have used the fact that $B_N = 0$ and
\begin{equation}
    \frac{\Var[w^p_s(X_1)]}{(\E[w^p_s(X_1)])^2} = \frac{\E[w^{2p}_s(X_1)]}{(\E[w^p_s(X_1)])^2} - 1 = \norm{\frac{\mu_s}{\mu_{s-1}}}_{L_{2p}(\mu_{s-1})} - 1
\end{equation}
On the other hand, applying Bernsetin's inequality  to $-w_s(X^i_{s-1}) < 0$ we obtain by a similar argument
\begin{equation}
    \Pr\Big(\frac{1}{N}\sum^N_{i=1} w^p_s(X^i_{s-1})  \leq (1 - \epsilon) \cdot \mu_{s-1}(w^p_s)  \Big) \leq \exp\Big\{-\frac{N \epsilon^2}{2 \norm{\frac{\mu_s}{\mu_{s-1}}}_{L_{2p}(\mu_{s-1})}} \Big\}
\end{equation}
Taking a union bound gives the result.
\end{proof}
Recall that the effective sample size (ESS) is defined as
\begin{align*}
    {E}_s := %\frac{
    \Big(\frac{1}{N}\sum^N_{i=1} w_s(X_i)\Big)^2
    %}{
    /
    \frac{1}{N}\sum^N_{i=1} w^2_s(X_i)
    %}
\end{align*}
Lemma~\ref{Appendix E: weight bound} can be used to bound ${E}_s$
\begin{lemma}
\label{Appendix E: ESS bound}
Suppose $0 < w_s(X^i_{s-1}) < 1$ and $X^{1:N}_{s-1} \sim \mu_{s-1}$ are independent and identically distributed. Then with probability $1 - \delta$,%\scsfootnote{Should be $\epsilon$s?  (I think lhs and rhs of inequalities should involve $\epsilon$'s, and this prob should remain $1-\delta$.)}
\begin{equation}
     \frac{(1-\epsilon)^2}{1+\epsilon} \leq {E}_s \cdot \norm{\frac{\mu_s}{\mu_{s-1}}}_{L_2(\mu_{s-1})}  \leq \frac{(1+\epsilon)^2}{1-\epsilon},
\end{equation}
%\scsfootnotetext{Should be flipped? (According to Eqn~(\ref{Eqn:TmpEqNo}) below, this should be ${E}_s \cdot \norm{\frac{\mu_s}{\mu_{s-1}}}_{L_2(\mu_{s-1})}$ ?  But I think it should be a bound on relative error?)}
for $N > \frac{2}{\epsilon^2} \cdot \log\left(\frac{4}{\delta}\right) \cdot  \norm{\frac{\mu_s}{\mu_{s-1}}}_{L_{4}(\mu_{s-1})}$
\end{lemma}

\begin{proof}
This follows by applying Lemma~\ref{Appendix E: weight bound} for $p = 1,2$ and taking a union bound.
\end{proof}

\section{Bounds on $L_2$ Distance}\label{sec:Monotonicity}

\begin{lemma}
\label{Appendix E: lp bound}
Suppose $\pi(x) \propto f(x)$, where $f(x)$ is a $d$-dimensional log-concave function. Let $\mu_s(x) \propto [f(x)]^{\beta_s}$. Then
\begin{align*}
    \norm{\frac{\mu_s}{\mu_{s-1}}}_{L_p(\mu_{s-1})} \leq  \left[\frac{\beta_s^p}{(p\beta_s - (p-1)\beta_{s-1})\beta^{p-1}_{s-1}} \right]^d \text{ for } p \geq 1
\end{align*}
\end{lemma}

\begin{proof} 
Let $Z(\beta) = \beta^d \int_{\mathbb{R}^d}[f(x)]^{\beta} dx$ and notice
\begin{align*}
\norm{\frac{\mu_s}{\mu_{s-1}}}_{L_p(\mu_{s-1})} &= \left[\frac{\beta_s^p}{(p\beta_s - (p-1)\beta_{s-1})\beta^{p-1}_{s-1}} \right]^d \\
&\times\frac{\left[Z(p\beta_s - (p-1)\beta_{s-1})\right]\left[Z(\beta_{s-1}) \right]^{p-1}}{\left[Z(\beta_s) \right]^p}
\end{align*}
Note that
\begin{align*}
\frac{1}{p}\left(p\beta_s - (p-1)\beta_{s-1}\right) + \left(1-\frac{1}{p}\right)\beta_{s-1} = \beta_s
\end{align*}
Lovasz and Vempala~\cite{LovaszGeometry} showed that $Z(\beta)$ is a log-concave function in $\beta$ and so $Z(\beta)^{\lambda}Z(\beta^{\prime})^{1-\lambda} \leq Z(\lambda \beta + (1-\lambda)\beta^{\prime})$ for $\lambda \in (0,1)$. Hence, setting $\lambda = \frac{1}{p}$ we have
\begin{align*}
\left[Z(p\beta_s - (p-1)\beta_{s-1})\right]^{\frac{1}{p}}\left[Z(\beta_{s-1}) \right]^{1-\frac{1}{p}} \leq Z(\beta_s)
\end{align*}
Taking the $p$th power on either side, it follows that
\begin{align*}
\frac{\left[Z(p\beta_s - (p-1)\beta_{s-1})\right]\left[Z(\beta_{s-1}) \right]^{p-1}}{\left[Z(\beta_s) \right]^p} \leq 1
\end{align*}
\end{proof}

\noindent If we consider an adaptive temperature selection scheme, we have an ordering on the size of $N$ needed to concentrate the ESS using Lemma~\ref{Appendix E: ESS bound} if we consider an adaptive temperature selection scheme. %Note that the 
The following result holds for \textit{any} tempered distribution. 
\begin{lemma}\label{Appendix E: monotone}
Suppose $\pi(x) \propto q_{\pi}(x)$ and $\mu_s(x) \propto q_{\mu_0}^{1-\beta_s}(x)q_{\pi}^{\beta_s}(x)$. Let $\beta^{\prime}_s = 2\beta_s - \beta_{s-1}$ and assume
\begin{align*}
    \frac{d}{d\beta_s} \int_{\X} q_{\mu_0}^{1-\beta^{\prime}_s}(x)q_{\pi}^{\beta^{\prime}_s}(x) dx =  \int_{\X}  \frac{d}{d\beta_s} q_{\mu_0}^{1-\beta^{\prime}_s}(x)q_{\pi}^{\beta^{\prime}_s}(x) dx, 
\end{align*}
%Then
\begin{align*}
\text{ Then } \qquad   \frac{d}{d\beta_s}\norm{\frac{\mu_s}{\mu_{s-1}}}_{L_2(\mu_{s-1})} = \frac{2 \norm{\frac{\mu_s}{\mu_{s-1}}}_{L_2(\mu_{s-1})}}{\beta_s - \beta_{s-1}}\left(D_{KL}(\mu_{s^{\prime}} || \mu_s) + D_{KL}(\mu_s || \mu_{s^{\prime}}) \right),
\end{align*}
where $\mu_{s^{\prime}}(x) = \frac{q_{\mu_0}^{1-\beta_s}(x)q_{\pi}^{\beta_s}(x)}{\int_{\X}q_{\mu_0}^{1-\beta_s}(x)q_{\pi}^{\beta_s}(x) dx}$ and $D_{\text{KL}}(P || Q)$ denotes the Kullback-Leibler divergence between distributions $P$ and $Q$.
\end{lemma}
Lemma~\ref{Appendix E: monotone} implies that the $L_2(\mu_{s-1})$ norm of $\mu_s / \mu_{s-1}$ is increasing in $\beta_s$ over the interval $(\beta_{s-1},1]$ for geometric path sequences schemes since $D_{\text{KL}}(P || Q) \geq 0$ for all $P$ and $Q$.

\begin{proof}
We let $h(x) = q_{\pi}(x) / q_{\mu_0}(x)$ and $Z(\beta) = \int_{\X} q_{\pi}^{\beta}(x)q_{\mu_0}^{1-\beta}(x) dx = \int_{\X} q_{\mu_0}(x) h^{\beta}(x) dx$. Notice
\begin{align*}
    \frac{d}{d\beta_s} Z(\beta^{\prime}_s) %&
    = 2  \int_{\X}q_{\mu_0}(x) \log(h(x)) h^{\beta^{\prime}_s}(x) dx, \ \ 
    \frac{d}{d\beta_s} Z^2(\beta_s) %&
    = 2 Z(\beta_s) \int_{\X} q_{\mu_0}(x)\log(h(x)) h^{\beta_s}(x) dx 
\end{align*}
Taking the derivative yields
\begin{align*}
    &\frac{d}{d\beta_s}\norm{\frac{\mu_s}{\mu_{s-1}}}^2_{L_2(\mu_{s-1})} %\\
    = %&
    \frac{d}{d\beta_s}\frac{Z(\beta^{\prime}_s) Z(\beta_{s-1})}{Z^2(\beta_s)} \\
    = &2 Z(\beta_{s-1})\frac{( Z^2(\beta_s)  \int_{\X} q_{\mu_0}(x)\log(h(x)) h^{\beta^{\prime}_s}(x) dx - Z(\beta^{\prime}_s) Z(\beta_s) \int_{\X} q_{\mu_0}(x)\log(h(x)) h^{\beta_s}(x) dx )}{Z^4(\beta_s)} \\
    = &\frac{2Z(\beta_{s-1})Z(\beta^{\prime}_s)}{Z^2(\beta_s)} \left(\int_{\X} q_{\mu_0}(x)\log(h(x))\left[\frac{h^{\beta^{\prime}_s}(x)}{Z(\beta^{\prime}_s)} - \frac{h^{\beta_s}(x)}{Z(\beta_s)}\right] dx \right) \\
    =  &\frac{2 \norm{\frac{\mu_s}{\mu_{s-1}}}^2_{L_2(\mu_{s-1})}}{\beta_s - \beta_{s-1}} \cdot  \Big((\beta_s - \beta_{s-1}) \int_{\X} \log(q_{\pi}(x))\left[\mu_{s^{\prime}}(x) - \mu_s(x)\right]dx \Big)
\end{align*}
Observe that $\beta_s - \beta_{s-1} = \beta^{\prime}_s - \beta_s$ and
\begin{align*}
     (\beta^{\prime}_s - \beta_s)\log\left(h(x) \right) =
     \log\Big(\frac{q_{\pi}^{\beta^{\prime}_s}(x)q_{\mu_0}^{1-\beta^{\prime}_s}(x)}{q_{\pi}^{\beta_s}(x)q_{\mu_0}^{1-\beta_s}(x)} \Big)  &= \log\Big(\frac{\mu_{s^{\prime}}(x)}{\mu_s(x)} \Big) + c,
\end{align*}
where $c > 0$ is a constant. Therefore,
\begin{align*}
&(\beta^{\prime}_s - \beta_s) \int_{\X} \log\left(h(x) \right) \left[ \mu_{s^{\prime}}(x) - \mu_s(x)\right] dx = \int \log\Big(\frac{\mu_{s^{\prime}}(x)}{\mu_s(x)} \Big)\left[ \mu_{s^{\prime}}(x)- \mu_s(x) \right]dx 
\end{align*}
It follows that
\begin{align*}
    \int \log\Big(\frac{\mu_{s^{\prime}}(x)}{\mu_s(x)} \Big)\left[ \mu_{s^{\prime}}(x)- \mu_s(x) \right]dx   
    &= \int \log\Big(\frac{\mu_{s^{\prime}}(x)}{\mu_s(x)} \Big) \mu_{s^{\prime}}(x)dx  +  \int \log\Big(\frac{\mu_s(x)}{\mu_{s^{\prime}}(x)}\Big)\mu_s(x) dx  \\
    &= D_{\text{KL}}(\mu_{s^{\prime}} || \mu_s) + D_{\text{KL}}(\mu_s|| \mu_{s^{\prime}})
\end{align*}
\end{proof}

\end{document}